\documentclass[12pt,epsf]{article}

\usepackage{graphicx}
\begin{document}
\baselineskip=0.7cm
\newcommand{\EQ}{\begin{equation}}
\newcommand{\EN}{\end{equation}}
\newcommand{\EQA}{\begin{eqnarray}}
\newcommand{\EQN}{\end{eqnarray}}
\newcommand{\EQAN}{\begin{eqnarray*}}
\newcommand{\EQNN}{\end{eqnarray*}}
\newcommand{\e}{{\rm e}}
\newcommand{\Sp}{{\rm Sp}}
\renewcommand{\theequation}{\arabic{section}.\arabic{equation}}
\newcommand{\Tr}{{\rm Tr}}
\newcommand{\lpartial}{\buildrel \leftarrow \over \partial}
\newcommand{\rpartial}{\buildrel \rightarrow \over 
\partial}
\renewcommand{\thesection}{\arabic{section}.}
\renewcommand{\thesubsection}{\arabic{section}.\arabic{subsection}}
\makeatletter
\def\section{\@startsection{section}{1}{\z@}{-3.5ex plus -1ex minus 
 -.2ex}{2.3ex plus .2ex}{\large}} 
\def\subsection{\@startsection{subsection}{2}{\z@}{-3.25ex plus -1ex minus 
 -.2ex}{1.5ex plus .2ex}{\normalsize\it}}
\makeatother
\makeatletter
\def\lesim{\mathrel{\mathpalette\gl@align<}}
\def\gtsim{\mathrel{\mathpalette\gl@align>}}
\def\gl@align#1#2{\lower.7ex\vbox{\baselineskip\z@skip\lineskip.2ex%
  \ialign{$\m@th#1\hfil##\hfil$\crcr#2\crcr\sim\crcr}}}
\makeatother

\newcommand{\Nt}{\widetilde{N}}
\newcommand{\bra}[1]{\langle #1\vert}
\newcommand{\ket}[1]{\vert #1\rangle}
\newcommand{\braket}[2]{\langle #1\vert #2\rangle}
\newcommand{\bbbk}[4]{{}_1\langle #1|{}_2\langle #2|
                      {}_3\langle #3|#4\rangle_{123}}
\newcommand{\C}{C_{\rm vac}}
\newcommand{\HI}{H_{\rm 3}}
\newcommand{\E}{E}
\renewcommand{\P}{{\bf P}}
\newcommand{\PSV}{P_{SV{\rm -}PS}}

\newcommand{\HSV}{\ket{\HI}_{SV{\rm -}PS}}
\newcommand{\HD}{\ket{\HI}_{DPPRT}}
\newcommand{\HCK}{\ket{\HI}_{CK}}

\newcommand{\vac}{{\rm vac}}
\renewcommand{\v}{{\rm v}}

\renewcommand{\a}{\alpha}
\renewcommand{\b}{\beta}
\newcommand{\inI}{2}
\newcommand{\inII}{3}
\newcommand{\out}{1}
\newcommand{\ainI}{\a_{(\inI)}}
\newcommand{\ainII}{\a_{(\inII)}}
\newcommand{\aout}{\a_{(\out)}}
\newcommand{\ar}{\a_{(r)}}
\newcommand{\as}{\a_{(s)}}
\newcommand{\asp}{\a_{(s')}}
\newcommand{\arp}{\a_{(r')}}
\newcommand{\atp}{\a_{(t')}}
\newcommand{\XI}{X_{\rm I}}
\newcommand{\XII}{X_{\rm II}}
\newcommand{\YI}{Y}
\newcommand{\yI}{Y_{\rm I}^{(1)}}
\newcommand{\YII}{Z}
\newcommand{\aDim}{\Delta^{\langle1\rangle}}
\newcommand{\Ctree}{C^{\langle0\rangle}}
\def\thefootnote{\fnsymbol{footnote}}

\begin{flushright}
hep-th/0409058\\
UT-KOMABA/04-10\\
September 2004
\end{flushright}
\vspace{0.3cm}
\begin{center}
\Large Impurity Non-Preserving 3-Point Correlators of BMN Operators \\
from PP-Wave Holography I : Bosonic Excitations

\vspace{0.7cm}

\normalsize
 \vspace{0.4cm}
Suguru {\sc Dobashi}
\footnote{e-mail address:\ \ {\tt doba@hep1.c.u-tokyo.ac.jp}}
and 
Tamiaki {\sc  Yoneya}
\footnote{
e-mail address:\ \ {\tt tam@hep1.c.u-tokyo.ac.jp}}

\vspace{0.3cm}

{\it Institute of Physics, University of Tokyo\\
Komaba, Meguro-ku, Tokyo 153-8902}

\vspace{1cm}
Abstract
\end{center}
\vspace{0.4cm}
As a continuation of our previous works studying the 
holographic principle in the plane-wave limit, we 
discuss the 3-point correlation functions of 
BMN operators  with bosonic excitations when  impurities 
are not conserved.  We show that our  proposal 
for a holographic mapping between the conformal 
OPE coefficients of super Yang-Mills theory 
and the 3-point vertex of 
the holographic string field theory is valid 
to the leading order in the large $\mu$ limit. 
Our results provide for the first time a direct holographic 
relation for the 3-point correlators of BMN operators 
including impurity non-preserving processes.

\newpage
\section{Introduction}
What is the correct interpretation of 
holographic principle  in the PP-wave limit of the 
AdS$_5$/SYM$_4$ correspondence has been one among 
several open problems since the 
first original discussion \cite{bmn} of the BMN limit. 
In the previous paper \cite{dy}, we proposed a 
simple direct relation between the 3-point 
OPE coefficients of conformal BMN operators 
and the bulk 3-point interaction vertex of string 
field theory, by developing the 
basic ideas presented first in ref. \cite{dsy}. 
In \cite{dy}, we constructed the `holographic' string 
field theory which meets requirements 
for the validity of the GKP-Witten relation and 
confirmed explicitly that it indeed reproduces the 3-point 
OPE coefficients of ${\cal N}=4$ super Yang-Mills theory 
in the leading large-$\mu$ expansion for stringy 
BMN operators in the impurity-preserving sector.  
Furthermore, it was also clarified that two entirely 
different proposals for relating 3-point vertices 
of (different versions of) string field theories to corresponding 
quantities on the gauge-theory side in 
the impuritiy-preserving sector are actually 
compatible to each other through our holographic 
map. 

It is important to recall, as we have stressed there,  that 
our holographic  relation in principle should be valid for 
much  more general impurity {\it non}-preserving 
processes too, 
since by construction the holographic relation is satisfied for the so-called `non-extremal'  
 3-point functions of chiral primary operators in the BMN 
limit. The latter functions are regarded to be the supergravity sector of general impurity non-preserving 3-point functions. 
The goal of the present paper is to confirm that 
our stringy holographic relation is indeed satisfied for 
impurity non-preserving cases   
in the leading large-$\mu$ expansion, which is  at 
the same level as  in the existing treatments of the 3-point functions of impurity-preserving sector on both sides. 

The significance of this result seems evident, 
since the usual discussions in most of other works being focused almost exclusively on the dilatation 
operator are practically 
restricted to impurity-preserving sector and therefore are 
difficult to extend them to impurity non-preserving 
processes.\footnote{
On the gauge-theory side, 
the matrix elements of a dilatation operator in the sense of 
renormalized perturbation theory 
would vanish unless the states have degenerate conformal dimensions 
in the lowest order. This is borne out in computations in 
refs. \cite{Beisertetal}\cite{Constable et. al. 1, Constableetal2}. 
} In contrast to this, the string-field Hamiltonian 
in our approach cannot {\it exactly} be interpreted 
  as the dilatation operator, since 
it generates an infinitesimal translation along a geodesic connecting from boundary to boundary in the Euclidean 
AdS spacetime. 
Except for regions approaching asymptotically  to 
the conformal boundary, such a translation cannot be identified with dilatation. 
In fact, this discrepancy between bulk Hamiltonian and 
 dilatation operator on the boundary is, in our opinion,  even more so 
in the usual derivation of the Penrose limit, leading  
to a geodesic which is disconnected from the boundary: 
There would be no asymptotic region where the Hamiltonian 
is related to the dilatation, corresponding to the fact that 
the translation with respect to the global time coordinate 
of the AdS spacetime does {\it not} coincide with  dilatation on the boundary at least within the usual geometric 
interpretation of the AdS spacetimes. This puzzle was  our main motivation for undertaking the present series of 
works \cite{dsy, yone, asy}, 
to which we would like to invite the reader for 
 further discussions of this basic issue.  From our viewpoint, 
the usual Penrose limit is obtained formally 
by a Wick rotation of the affine time parameter along the 
geodesics connecting boundary to boundary. 
Note  that the relative topology of 
trajectory with respect to the conformal boundary 
is completely changed by this Wick rotation. 
This however 
explains why the dilatation comes about even 
in Minkowski formulation. In any case, 
given the apparent correspondence 
for conformal dimensions $\{\Delta_r\}$ on one hand, 
it would be very strange if there could be no clear way of mapping the OPE coefficients which,  together with conformal dimensions, constitute the crucial 
data  of CFT. 

We  note that, even though the matrix elements of 3-point 
interaction vertices of string field theory in the
 impurity non-preserving sector 
 often have additional 
powers in $1/\mu$,\footnote{
This is due to an additional power of $1/\mu$ in the 
asymptotic forms of Neumann functions other than 
$\tilde{N}^{21}_{mn}$ and $\tilde{N}^{31}_{mn}$. See the Appendix.
} the OPE coefficents are not 
necessarily  of lower order in the large-$\mu$ expansion 
for {\it large but  fixed} $R$-charge angular momentum $J$.  
It is also well known that a class of impurity non-preserving 
processes cannot simply be ignored in summing up 
intermediate states in higher-loop calculations, at least  
within the logic of string field theory as presently 
understood. Once the summation over 
intermediate states are involved, the limiting 
procedures required in the 
studies of amplitudes for BMN states should be a very subtle problem, since results would depend on the orders 
of various infinite sums and the limiting procedures. 
We hope that our concrete results  
for impurity non-preserving 3-point 
correlators provide a useful first step for further 
exploration of the AdS holography in the plane-wave limit 
beyond the approximation of keeping only the 
impurity preserving sector.  

In the next section, we start from summarizing briefly the 
holographic relation and the holographic 
string field theory proposed in the previous work. 
In section 3, we treat a general case  of impurity 
non-preserving 3-point functions 
where only non-singlet (more precisely, traceless) 
scalar (and bosonic) 
impurities corresponding to the fluctuations 
along 4-directions among the $S^5$ of AdS$_5\times S^5$ 
are involved. 
In section 4, we extend the results to include non-singlet 
vector excitations. In section 5, we discuss a few typical cases  
where singlet impurities are involved.   
 The concluding 
section 6 contains some remarks
on relevant issues, especially, the uniqueness of 
our holographic relation and the question of 
higher-order effects. Some useful formulas which are 
relevant for discussions in the main text 
are summarized in the Appendix. 

\section{The holographic relation and string 
field theory} 

The key concept behind our holographic relation is 
that the correspondence between conformal boundary 
and bulk spacetime in the plane-wave limit of the 
AdS/CFT correspondence should be based on a
 tunneling geodesics traversing inside the AdS spacetime 
with the Euclidean signature which start 
from the conformal boundary and end again at 
conformal boundary, 
instead of the usual procedure of taking the Penrose 
limit in the AdS space-time with Minkowski signature. 
Such amplitudes can be regarded as an Euclidean analog of 
the usual S-matrix. 
This picture, which has been previously 
proposed by us for the purpose of  resolving 
 some puzzles associated with 
holography in the PP-wave limit in  \cite{dsy},  
automatically emerges by studying the large-$J$ limit of the 
GKP-Witten relation for single-trace 
chiral primary operators. 
Since the 3-point functions of chiral primary operators 
are protected against the perturbative corrections 
on the gauge-theory side, we can derive the effective action 
describing the interaction of the BMN operators along the 
tunneling geodesics in the supergravity sector. 
The full string field theory, the `holographic' string field 
theory, should reduce to this effective action 
in the supergravity sector. 

Under this requirement, 
we are led to a unique 3-point vertex 
which should obey the holographic relation 
based on the above tunneling picture. 
We stress that in the GKP-Witten conjecture the bulk partition function as a 
functional of boundary fields is fixed uniquely with 
a given basis of linearized supergravity fields, 
independently of field redefinitions. For 3-point 
correlators, the effect of such a field redefinition would be 
proportional to the equation of motion, and  integration 
over the interaction point in the bulk would give vanishing result.  In this sense, even though 
`energies' being now the conformal dimensions are not conserved, the Euclidean S-matrix is very similar \cite{dsy} 
to the ordinary S-matrix of flat Minkowski spacetime. 
We will give a remark on the issue of uniqueness of our 
prescription in the final section of the present paper. 
A subtlety in the case of the extremal case where 
the conformal dimensions are strictly conserved  
will also be clarified.

The holographic relation is summarized as 
\begin{equation}
C_{123}=
{\tilde{\lambda}_{123}
\over \mu(\Delta_2+\Delta_3-\Delta_1)} , 
\label{clambdarelation1}
\end{equation}
\begin{equation}
\tilde{\lambda}_{123}
=\Big(f {J_2J_3\over J_1}\Big)^{-(\Delta_2+\Delta_3-\Delta_1)/2}\Gamma({\Delta_2+\Delta_3-\Delta_1 \over 2}+1)
\lambda_{123}
\label{clambdarelation2}
\end{equation}
with 
\begin{equation}
\lambda_{123}
=\,  _{(1)}\langle 
1|   _{(2)}\langle 2|  _{(3)}\langle 3| 
{\sqrt{J_1J_2J_3}\over N}\, 
\, |H_3\rangle_h , 
\label{clambdarelation3}
\end{equation}

The conformal dimensions $\Delta_r$ are 
those in the planar limit. 
The CFT coefficients $C_{123}$ is defined under the 
standard normalization of two point functions 
as
 \EQ
\langle \overline{O}_{1}(\vec{x}_1)O_{2}(\vec{x}_2)\rangle =
{\delta_{12} \over |\vec{x}_{12}|^{2\Delta_1}}
\EN 
by
\EQ
\langle \overline{O}_{1}(\vec{x}_1) O_{2}(\vec{x}_2) 
O_{3}(\vec{x}_3) \rangle
={C_{123}
\over |\vec{x}_{12}|^{2\alpha_3}|\vec{x}_{23}|^{2\alpha_1}
|\vec{x}_{31}|^{2\alpha_2}}
\label{gene3pointcorr}
\EN
with ($\vec{x}_{12}=\vec{x}_1-\vec{x}_2, \, etc$)
\EQ
\alpha_1 ={\Delta_2 + \Delta_3 -\Delta_1 \over 2} ,\quad etc.
\label{alpha1}
\EN
The symbol $|H_3\rangle_h$ is the 
3-point interaction vertex of the holographic string field theory  
\begin{equation}
S_3=
{1\over 2}\int d\tau  \,  _{(1)}\langle 
\overline{\psi}|   _{(2)}\langle \psi|  _{(3)}\langle \psi| {\sqrt{J_1J_2J_3}\over N}|
H_3\rangle_h  
+ h.c. \, .
\label{s3zero1}
\end{equation}
Here the integration over the $R$-charge angular momenta 
$J_r$ associated with the string fields 
$_{(r)}\langle \psi| \, \, (r=2,3), \, 
_{(1)}\langle \overline{\psi}| $ 
 should be implicitly understood, under the conservation condition $J_1=J_2 +J_3$. 
The quantity $f$ in (\ref{clambdarelation2}) depending on  
$\alpha'$ is responsible for the 
nonlocality caused by the extended nature of strings and is given by 
\EQ
f=1-4\mu\alpha_{(1)}\alpha_{(2)}\alpha_{(3)} K
\EN
where $K$ is the well known expression in string 
field theory which is defined in terms of 
various infinite matrices associated with the familiar 
overlap condition for string interaction. 
For full details of the 
string field theory, we refer the reader to the Appendix B 
of the previous paper \cite{dy} and to the references cited therein and below. 
The 3-point interaction vertex takes just an equal-weight sum 
of the two previously constructed vertices which are 
compatible with supersymmery algebra:
\EQ
|H_3\rangle_h\equiv {1\over 2}(|H_3\rangle_{SV}+|H_3\rangle_D)
\EN
where  $|H_3\rangle_{SV}$ and $|H_3\rangle_D$ are 
those proposed in \cite{SV} 
and in \cite{DPPRT}, respectively. In particular, the prefactor of 
the part $|H_3\rangle_D$ manifestly takes the form of the energy difference
$2\mu \alpha_1$ while that of the part $|H_3\rangle_{SV}$ is 
the one obtained by a natural generalization of the familiar 
flat-space vertex. 

Our convention 
for various parameters is as follows:  
$R^4/J_1^2(\alpha')^2 =\lambda'=1/(\mu p_1^+ \alpha')^2=
g_{{YM}}^2N/J_1^2, g_2=J_1^2/N$,  $\alpha_{(r)}=\alpha'p^+_r$, 
$\mu |\alpha_{(r)}|=\alpha'J_r/R^2$ $\, (r=1,2,3)$. 
Do not confuse the symbol $\alpha_{(1)}, \, etc $ with 
the previously defined $\alpha_1, \,  etc$ as 
(\ref{alpha1}). 
Using these conventions, the large $\mu$ limit 
of $f$ is given as 
\EQ
f{J_2J_3\over J_1} \rightarrow {J_1\over 4\pi \mu |\alpha_{(1)}|}\, .
\label{asympf}
\EN
We assume $\alpha_{(1)} (=-\alpha_{(2)}-\alpha_{(3)}) <0, \alpha_{(2)}>0, \alpha_{(3)} 
>0$.  Note that 
the mass parameter $\mu$ becomes a meaningful curvature 
parameter $\mu=1/R$ as we 
identify the light-like momentum with 
the angular momentum by $|p^+_r| = J_r/R$, which is 
the correct one for comparing the action with flat-space form. 
The interaction term is of order $g_s$ for fixed 
$R$ and $p^+$ as it should be. 
The overall factor $\sqrt{J_1J_2J_3}/N$ is rewritten as 
$4\pi g_s(\alpha')^2\sqrt{|p_1^+p_2^+p_3^+|}/R^{5/2}$
in terms of the usual string coupling constant $g_s$.  
This form indicates that the effective action 
in fact has a nontrivial curvature dependence.

The above normalization of the interaction vertex is 
fixed by matching the effective action 
in the supergravity sector.  The string field is normalized such that the free action takes the standard {\it Euclidean} form 
\EQ
S_{2}=\int d\tau \Big[{1\over 2}(\langle \overline{\psi}| \partial_{\tau}|\psi\rangle 
-{1\over 2}(\partial_{\tau}\langle \overline{\psi}|)|\psi\rangle +
\langle \overline{\psi}| H_2 |\psi\rangle \Big] , 
\EN
\begin{equation}
H_2^{(r)}=\frac{1}{|\alpha_{(r)}|}\sum_{n=-\infty}^{\infty}
\omega_n^{(r)}(a_n^{(r)\dagger}a_n^{(r) }
+b_n^{(r)\dagger}b_n^{(r)}) \quad \mbox{with} \quad  \omega^{(r)}_n=
\sqrt{n^2 + (\mu\alpha_{(r)})^2} ,
\end{equation}
where $(a_n^{(r)}, a_n^{(r)\dagger})$'s and $(b_n^{(r)}, 
b_n^{(r)\dagger})$'s  represent 8-component 
bosonic and fermionic oscillators of $r$-th string. 

One of the most 
characteristic features of this string field theory 
is that for purely bosonic string states the (total) 
prefactor reduces to the special form which consists 
of only $\cos$ modes $a_n^{(r)}\,  (n \ge 0$) \cite{CK} and of only $\sin$ modes $a_{-n}^{(r)} \, (n >0)$ 
for scalar ($i=5 \sim 8$) and vector ($i=1\sim 4$) impurities, respectively : 
 \EQ
|H_3\rangle_h \Rightarrow 
\sum_{r=1}^3
\left(\sum_{i=5}^8\sum_{m=0}^{\infty}
\frac{\omega^{(r)}_{m}}{\alpha_{(r)}}
a^{(r)i\dagger}_{m}a^{(r)i}_{m}
+\sum_{i=1}^4\sum_{m=1}^{\infty}
\frac{\omega^{(r)}_{m}}{\alpha_{(r)}}
a^{(r)i\dagger}_{-m}a^{(r)i}_{-m}
\right)\ket{\E_a} ,
\label{bosonicpart}
\EN
where $|E_a\rangle$ given as 
\EQ
 \hspace{-1cm}\ket{E_a}=\exp
 \left[
 -\frac12 
 \sum_{r,s=1}^{3}\sum_{m,n=-\infty}^{\infty}\alpha_m^{(r)\dagger} \tilde{N}_{mn}^{rs} \alpha_n^{(s)\dagger}
 \right]\ket{0}_{(1)(2)(3)}
\EN
is the standard bosonic overlap vertex.\footnote{
As in \cite{dy}, the minus sign on the exponential 
factor is due to our phase convention in defining 
the world-sheet oscillators. This is different from the 
standard one in the literature, but is necessary for 
matching between bulk and boundary. 
}
Thus, $Z_2$ (or higher SO(8) symmetry) of the 
free string-field theory is completely violated by the 
interaction. 
The oscillators ($\alpha_n^{(r)}, \alpha_n^{(r)\dagger}, 
\, n\in {\bf Z}$)  
in the exponential basis which directly 
corresponds to the standard convention for the 
BMN operators are related to the trigonometric basis by  ($n> 0$)
\begin{eqnarray}\label{sin/cos-exponential}
 \a_0=a_0,\quad
\a_{n}=\frac{1}{\sqrt{2}}(a_n-ia_{-n}),\quad
\a_{-n}=\frac{1}{\sqrt{2}}(a_n+ia_{-n}). 
\end{eqnarray}

In our formalism, the BMN operators 
must have  definite conformal dimensions  to at least  
the first order in $g_2$ and to all orders with 
respect to $\lambda'$. 
This means that for unprotected stringy BMN operators 
we have to take into 
account the effect of various 
operator mixing including double-trace 
operators \cite{Beisertetal} \cite{Constableetal2} to the 
leading approximation in $1/N$ expansion in extracting 
the CFT coefficients $C_{123}$ on the gauge-theory side.  
In real life, we have to be 
satisfied by studies of leading $1/\mu$-expansions at 
current stage of development, since computation of 3-point OPE 
coefficients at  the order $g_2\lambda'$ and beyond has not been carried out so far. Such a computation on the gauge-theory side in general 
requires two-loop calculations including operator mixing 
effects. 

In \cite{dy}, we have 
presented a general argument that the above relation must be 
correct for arbitrary impurity-preserving 3-point functions,  
by reinterpreting appropriately the previously known results \cite{gomisetal} 
\cite{DPPRT} for comparison 
between string field theory vertices and gauge-theory 
calculations in the leading large-$\mu$ expansion. 
We explicitly confirmed the above relation for two-impurity processes including vector and spinor excitations . 
The peculiar correction factor $\Big(f {J_2J_3\over J_1}\Big)^{-(\Delta_2+\Delta_3-\Delta_1)/2}\Gamma({\Delta_2+\Delta_3-\Delta_1 \over 2}+1)
$ appearing in the 
expression (\ref{clambdarelation3}) can be neglected 
 to the leading order in the $1/\mu$-expansion for 
impurity-preserving processes.  
Thus, in these cases, the relation reduces to the 
one first conjectured in ref. \cite{Constable et. al. 1} 
from a different viewpoint.  
Our previous work, however, 
 clarified that this particular relation  
is valid only with our holographic string field theory vertex 
$|H_3\rangle_h$ which takes into account  
the operator mixing of gauge-theory operators 
by the above specific combination of two 
different prefactors.  

When the impurities are not conserved, 
the correction factor plays a crucial role, as 
has already been shown by the construction 
in \cite{dy} for non-extremal correlators of chiral primary 
operators on the basis of the GKP-Witten relation. 
Our task is now to confirm this for unprotected stringy BMN 
operators by studying the large-$\mu$ limit.  
We can divide the impurity non-preserving 
3-point interactions into two classes, class I and II, 
respectively, depending on 
$\alpha^{cl}_1=(\Delta^{cl}_2+\Delta_3^{cl}-
\Delta_1^{cl})/2>0$ or  $<0$ where 
$\Delta^{cl}_r$ denotes the classical conformal 
dimension, counting the number of fields and (spacetime) derivatives 
involved in each BMN operator. 
Because of the SO(4) $\times$ SO(4) symmetry, 
class II processes are possible only when 
we allow singlet representations for the external 
line 1. 

\section{Class I non-singlet scalar impurities}
\setcounter{equation}{0}
Let us start from considering 
a simple  example of class I  processes. 
We denote the directions  
of scalar excitations by $i, j, \ldots \in (5\sim 8)$. 
The operator 2 is assumed to involve 4 scalar excitations 
in all different directions $i, j, k, \ell$ with world-sheet 
momenta $m, -m, p, -p,$ respectively. The operators 3 and 1 
are assumed to involve 2 scalar impurities in directions 
$k,   \ell$ with momenta $q, -q$ and directions 
$i, j$ with momenta $n,-n$, respectively. The explicit 
forms are given, suppressing possible mixing terms 
with double-trace operators, as 
\EQA
O^{(2)}_{(i,m;j,-m;k,p;\ell,-p)}=
{1\over \sqrt{(J_2+3)^3 N^{J_2+4}}}\sum_{a+b+c+d=J_2}&&\hspace{-1cm}
\Tr
\Big[
\phi_iZ^a\phi_jZ^b\phi_kZ^c\phi_{\ell}Z^d \nonumber \\ 
\times \quad \e^{{2\pi i\over (J_2+3)}[-am + (a+b+1)p -(a+b+c+2)p]}
   && + \quad \mbox{ permutations} \, 
\Big]
\EQN 
where the `permutations' indicates  the summation 
over all non-equivalent positioning of the impurities 
($a, b, c, d \ge 0$), and 
\EQA
O^{(3)}_{(k,q;\ell, -q)}&=&{1\over \sqrt{(J_3+1) N^{J_3+2}}}
\sum_{a=0}^{J_3}\Tr
\Big[
\phi_k Z^a \phi_{\ell}Z^{J_3-a}\, \e^{-{2\pi i\over (J_3+1)}aq}
\Big] , \\
O^{(1)}_{(i,n;j, -n)}&=&{1\over \sqrt{(J_1+1) N^{J_1+2}}}
\sum_{a=0}^{J_1}\Tr
\Big[
\phi_i Z^a \phi_jZ^{J_1-a}\, \e^{-{2\pi i\over (J_1+1)}an}
\Big].
\EQN
The overall constants correspond to 
  the normalization of scalar fields 
such that the free propagators  
are equal to  (no summation over $i$) $\langle \phi_i(x_1)\phi_i(x_2)\rangle 
=\langle \overline{Z}(x_1)Z(x_2)\rangle =1/
|x_{12}|^2 $.  Note that the phases associated with 
nonzero momenta are determined only by relative (and oriented) 
distances among impurities. Here we have adopted the 
normalization constants and phase factors which are 
slightly different from those adopted 
in the recent literature. In the large $J$ limit, 
the difference between $J$ and $J+k-1$ 
with $k$ being the number of impurities is 
 inconsequential for our leading order 
computations restricted to bosonic excitations. 
Therefore, in the following we will ignore 
this difference unless otherwise stated explicitly and 
use the more familiar convention by ignoring these shift 
of $J$, in order to save the space for mathematical 
expressions. 
However, for fermionic excitations it turns out that 
these shifts in the phases actually play an important role. 
Fermionic excitations will not, however, be treated in the 
present paper and be left for a separate work. 

It is obvious that the gauge-theory correlator $\langle \overline{O}^{(1)}_{(i,n;j, -n)}
O^{(2)}_{(i,m;j,-m;k,p;\ell,-p)}
O^{(3)}_{(k,q;\ell, -q)}\rangle$ in the leading 
planar approximation 
 must necessarily involve 
contractions of impurity fields $\phi_k$ and $\phi_{\ell}$ between 2 and 3, aside those between 1 $\leftrightarrow$ 2 and 
1 $\leftrightarrow$ 3. 
This implies that to the lowest nontrival order in $g_2$ 
we can ignore  mixing with double-trace 
operators in calculating the 3-point function, 
since the effect of the mixing terms  
becomes of higher order with respect to 
$1 \over N$, 
in contrast to the impurity preserving case 
where there is no contraction 
between 2 and 3 for the mixing contribution which 
corresponds to the topology of a product of two cylinders ($\times 1/N$). 
We then find by a straightforward free-field computation that the CFT coefficient is given by
\EQ
C_{123}={1\over N} {1\over \sqrt{J_1J_2^3 J_3}}
\times 2J_1^2 {\sin^2(\pi y n)\over \pi^2(n-{m\over y})^2}
= {\sqrt{J_1J_2J_3}\over N}{2\over  J_1^2\,  y^2(1-y)}
{\sin^2(\pi y n)\over \pi^2(n-{m\over y})^2}
\label{cftcoeffexample}
\EN
with $y=J_2/J_1$. The first factor in the first equality 
comes from the overall normalization, and the 
second factor is the result of free-field contractions and of 
summation over the positioning of impurities. In particular, 
the factor 2 originates from the permutation 
of two contractions between operators 2 and 3.  Since the 
contractions between 2 and 3 can occur in the planar 
limit only when the  
impurity fields $\phi_k$ and $\phi_{\ell}$ are 
adjacent to each other, they do not have any 
momentum dependence in the present large-$J$ limit. 
This is in general true 
for arbitrary configurations of momenta of such 
impurities at least for exchanges of bosonic 
fields, beyond the above special case. 

Now let us turn to the corresponding calculation 
in string field theory. The string states are 
\EQ
 _{(1)} \langle0|\alpha_n^{i (1)}
\alpha_{-n}^{j(1)}, \quad 
 _{(2)}\langle 0|\alpha_m^{i (2)}
\alpha_{-m}^{j(2)}\alpha_p^{k(2)}\alpha_{-p}^{\ell(2)}, 
\quad 
 _{(3)}\langle  0|\alpha_q^{k (3)}
\alpha_{-q}^{\ell(3)}, 
\EN
respectively. The relevant part of the 
interaction vertex takes the form
\[
|H_3\rangle_h \Rightarrow P_{123}\exp[ -{\cal N}^{12}  -{\cal N}^{23}]\, |0\rangle_{(1),(2),(3)}
\]
where the prefactor is 
\[
P_{123}={\omega_m^{(2)}\over \alpha_{(2)}}
(2+\alpha_m^{(2)\dagger}\alpha_{-m}^{(2)}+
\alpha_{-m}^{(2)\dagger}\alpha_{m}^{(2)})
+{\omega_p^{(2)}\over \alpha_{(2)}}
(2 +\alpha_p^{(2)\dagger}\alpha_{-p}^{(2)}+
\alpha_{-p}^{(2)\dagger}\alpha_{p}^{(2)})
\]
\EQ
+{\omega_q^{(3)}\over \alpha_{(3)}}
(2+\alpha_q^{(3)\dagger}\alpha_{-q}^{(3)}+
\alpha_{-q}^{(3)\dagger}\alpha_{q}^{(3)})
-{\omega_n^{(1)}\over |\alpha_{(1)}|}
(2+\alpha_n^{(1)\dagger}\alpha_{-n}^{(1)}+
\alpha_{-n}^{(1)\dagger}\alpha_{n}^{(1)})
\EN
and ${\cal N}^{rs}$ are expressed in terms of the 
Neumann functions in the exponential basis as 
\[
{\cal N}^{12}=\tilde{N}_{nm}^{12}(\alpha_n^{(1)\dagger}
\alpha_m^{(2)\dagger} +\alpha_{-n}^{(1)\dagger}
\alpha_{-m}^{(2)\dagger})
+\tilde{N}_{-nm}^{12}(\alpha_{-n}^{(1)\dagger}
\alpha_m^{(2)\dagger}+\alpha_n^{(1)\dagger}
\alpha_{-m}^{(2)\dagger})
\]
\EQ
\hspace{1cm} 
+\tilde{N}_{np}^{12}(\alpha_n^{(1)\dagger}
\alpha_p^{(2)\dagger} +\alpha_{-n}^{(1)\dagger}
\alpha_{-p}^{(2)\dagger})
+\tilde{N}_{-np}^{12}(\alpha_{-n}^{(1)\dagger}
\alpha_p^{(2)\dagger}+\alpha_n^{(1)\dagger}
\alpha_{-p}^{(2)\dagger}), 
\EN
\EQ
{\cal N}^{23}=\tilde{N}_{pq}^{23}(\alpha_p^{(2)\dagger}
\alpha_q^{(3)\dagger} +\alpha_{-p}^{(2)\dagger}
\alpha_{-q}^{(3)\dagger})
+\tilde{N}_{-pq}^{23}(\alpha_{-p}^{(2)\dagger}
\alpha_q^{(3)\dagger} +\alpha_{p}^{(2)\dagger}
\alpha_{-q}^{(3)\dagger}).
\EN
Note that there is no contraction between $1\leftrightarrow 3$. 

In the leading large $\mu$ limit we can simply replace 
the energy factor $\omega_n^{(r)}/|\alpha_{(r)}|$ by 
$\mu$ for  class I processes, since there is no 
singularity in the multiplying factor (denoted by $G$) 
of the holographic relation
\EQ
G\equiv 
\Big(f {J_2J_3\over J_1}\Big)^{-(\Delta_2+\Delta_3-\Delta_1)/2}
{\Gamma({\Delta_2+\Delta_3-\Delta_1 \over 2}+1)\over 
\mu(\Delta_2+\Delta_3-\Delta_1)}
\sim 
{\alpha_1^{cl}\, !\over 2\mu\alpha_1^{cl}}
\Big({J_1 \over 4\pi \mu|\alpha_{(1)}|}\Big)^{-\alpha_1^{cl}}, 
\EN
with $\alpha_1 = (\Delta_2+\Delta_3-\Delta_1)/2 
\sim \alpha_1^{cl} + O(1/\mu^2)$.  In the present example, 
$\alpha_1^{cl}=2$. 
Using the properties of Neumann functions such as 
$\tilde{N}_{nm}^{rs}=\tilde{N}_{-n-m}^{rs}, 
\tilde{N}_{-nm}^{rs}=\tilde{N}_{n-m}^{rs}$ for 
$m, n > 0$, we can easily see that the contributions from two 
parts of the prefactor involving $\alpha_m^{(2)\dagger}\alpha_{-m}^{(2)}+
\alpha_{-m}^{(2)\dagger}\alpha_{m}^{(2)}$ 
and $\alpha_n^{(1)\dagger}\alpha_{-n}^{(1)}+
\alpha_{-n}^{(1)\dagger}\alpha_{n}^{(1)}$, respectively,  
cancel against each other.  This cancellation 
corresponds to the fact that two terms 
$|H_3\rangle_{SV}$ and $|H_3\rangle_D$ 
have equal contributions in this case. 
The situation  is in contrast to the 
impurity-preserving sector where they play different 
roles (roughly speaking, `bare' interaction and 
mixing effect, respectively), owing to the existence of the 
singularity in the factor $G$. 
 Furthermore, 
using the property that the large $\mu$ limit of 
$|\tilde{N}_{pq}^{23}|$ is independent of 
the momenta, we find that 
the matrix element of the interaction vertex is 
equal to 
\[
4\mu\Big(\tilde{N}^{12}_{nm}\tilde{N}^{23}_{pq}
\Big)^2 .
\]
Using the explicit expressions  of Neumann functions 
(see the Appendix for a summary)  
in the leading large $\mu$ limit, 
\EQ
\tilde{N}^{12}_{nm}={(-1)^{m+n+1}\sin(\pi n y)\over 
\pi \sqrt{y}(n-{m\over y})}, 
\quad \tilde{N}^{23}_{pq}=
{(-1)^{p+1}\over 4\pi \mu |\alpha_{(1)}|\sqrt{y(1-y)}}\, , 
\EN  
we find that 
 the CFT coefficient (\ref{cftcoeffexample}) 
from gauge-theory side precisely matches the string 
interaction vertex with the 
holographic relation (\ref{clambdarelation1}) $\sim$
 (\ref{clambdarelation3}). 

It is not difficult to extend the above result to a more 
general case of scalar impurities. In this section, we  
restrict ourselves to operators without any singlet 
representation with respect to O(4) group 
of rotations of scalar directions.  
More precisely, the scalar impurities 
$\phi_{i_1}, \phi_{i_2}, \ldots$ 
are contracted to `polarization' tensors 
$C^{I}_{i_1i_2 \cdots }$ which are traceless 
with respect to arbitrary pair of the O(4) indices 
$(i_1, i_2, \ldots)$ for each conformal BMN operator 
in an appropriate irreducible representation of 
SO(4) :
\[
O^I_{(p_1, p_2, \ldots)}
= {C^I_{i_1i_2 \cdots }\over \sqrt{J^{k-1}N^{J+k}}}\sum_{a_1+a_2 + \cdots =J}
\hspace{-0.3cm} \Tr 
\Big[
\phi_{i_1}Z^{a_1}\phi_{i_2}Z^{a_2} 
\cdots 
\e^{2\pi i(a_1p_2 +a_2p_3 + \cdots)/J} 
\]
\EQ 
\hspace{3.5cm}+\quad \mbox{permutations} \quad \Big] , 
\label{genetracelessop}
\EN
where $k \, (\ge 1)$ is the number of impurities and 
as above we have to sum over non-equivalent positioning 
of impurities.\footnote{
Note that we have changed the notations slightly from 
ref. \cite{dy}. For instance, $\tilde{k} \rightarrow k$. 
The convention on the summation over `permutations' 
(for sugra modes in particular) is also 
changed. In the present paper, `non-equivalent' permutations 
exclude those which correspond to cyclic permutations, while in 
\cite{dy} they were not excluded and hence the normalization 
constants had an additional power of $1/J$. 
} The traceless condition allows 
us to ignore 
the mixing of the pairs of operators $Z$ and $\overline{Z}$ 
which would be needed \cite{PR} for defining operators 
with definite conformal dimensions even if the mixing with 
double-trace operators can be ignored.  If we consider higher 
orders in $g_{{\rm YM}}$, mixing between purely bosonic operators with antisymmetrized 
scalar indices (or mixed scalar and vector indices) and those involving fermion impurities 
are expected to play important roles. In this paper, we ignore such 
complications by restricting ourselves to the leading 
order effect in the large-$\mu$ expansion on 3-point functions.   The symbol $I$ of the 
polarization tensor indicates collectively  the configurations  
of both momenta and SO(4) representation. 
Its normalization is most conveniently expressed using 
oscillators $a_p^{(1)\dagger}, a_p^{(1)}, etc$ of strings, as 
\EQ
_{(1)}\langle C^{I_1}|\, _{(2)}\langle C^{I_2}| 
\exp\Big[\sum_{p=-\infty}^{\infty}\alpha_{p}^{(1)\dagger}\alpha_p^{(2)\dagger}
\Big]|0
\rangle_{(1)(2)} =\delta^{I_1I_2},
\EN
\EQ
_{(1)}\langle C^{I_1}|=
_{(1)}\langle 0|C^{I_1}_{i_1i_2\cdots}\alpha_{p_1}^{i_1(1)}
\alpha_{p_2}^{i_2(1)}
\cdots 
\quad etc, 
\EN
where the SO(4) indices on the exponential is suppressed. 
According to this convention, various symmetry factors associated with the 
symmetry property of the SO(4) indices are absorbed in 
the normalization of polarization tensors themselves. 
This representation naturally takes into account the summation 
over all allowed  
contractions  with equal weights 
corresponding to those of free-gauge theory.  
Of course, the momenta must satisfy 
the level matching condition
\EQ
\sum_{i=1}^k p_k =0. 
\EN

Suppose the $r$-th operator  has $k_r$ 
impurities. Then, the number of contractions between 
$r=2$ and $r=3$ is equal to 
\[
\alpha_1^{cl}=(k_2+k_3-k_1)/2. 
\]
Now, in computing the gauge-theory correlators for this general type 
of operators, we note 
the following properties, all of which already appeared in the above example. 
\begin{enumerate}
\item We can ignore operator 
mixing with double-trace operators, since the mixing contributions are 
always of higher order in the $1/N$-expansion because of the 
existence of nonzero number of 2-3 contractions. 
In the absence of the mixing with doubel-trace operators, the 
CFT coefficients to the leading order in 
$1/\mu$ can be determined entirely from the 
free-field contractions which give the correct 
leading order form of the spacetime factor of conformal 
3-point functions. 
\item In the planar limit, 
the contractions 
between 2 and 3 must belong to a single group of 
adjacent products of the impurity $\phi_i$ 
fields in which no $Z$ field is contained, since all 
$Z$ fields in the operators 2 and 3 must be contracted with $\overline{Z}$ 
fields in the operator 1 and hence would lead to a 
nonplanar contribution otherwise. 
In other words, in the sum over $a_i$ in the 
definition (\ref{genetracelessop}), such contractions  are 
 possible only when $a_{i_{s}}=a_{i_{s+1}}=\cdots =
a_{i_{s+\alpha_1-1}}=0$ with some $s$ for 
a consecutive set of impurities $(\phi_{i_s}, 
\phi_{i_{s+1}}, \ldots, \phi_{i_{s+\alpha_1-1}})$.  See fig. 1. 
\item By the cyclic symmetry of $\Tr$ operation and 
the level-matching condition, 
we can always choose $s=1$. This means that we can 
replace the momentum-dependent exponential by one 
for these impurities in the large-$J$ limit. 
Therefore, the $2\leftrightarrow 3$ 
contractions do not give any momentum dependent 
factor. 
\end{enumerate}

\begin{center}
\begin{figure}
\begin{picture}(100,100)
\put(160,0){
\includegraphics[width=100pt]{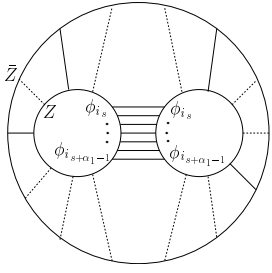}}
 \end{picture}
\caption{The dotted lines represent $Z$-$\overline{Z}$ 
contractions, while real lines represent contractions 
of scalar fields $\phi_i$'s.  The outer circle 
denotes the trace of operator product in the line 1 and two inner circles the traces of operator products in the line 2 and 3, respectively. 
}
\label{fig1} 
\end{figure}
\end{center}
Using these properties, 
 the 3-point correlation function is computed as follows. 
First, the overall constant factor is 
\[
\alpha_1^{cl}! {N^{J_1+k_1 + \alpha_1^{cl}-1} 
\over \sqrt{\prod_{i=1}^3J_i^{k_i-1}N^{J_i+k_i}
}}={\sqrt{J_1J_2J_3}\over N}
\alpha_1^{cl}! J_1^{-(k_1+k_2 +k_3)/2}
\Big({J_2\over J_1}\Big)^{-k_2/2}
\Big({J_3\over J_1}\Big)^{-k_3/2}
\] 
which can also be written as 
\EQ
{\sqrt{J_1J_2J_3}\over N}
 \, \times \, J_1^{-k_1}\Big({J_2\over J_1}\Big)^{-\overline{k}_2/2}
\Big({J_3\over J_1}\Big)^{-\overline{k}_3/2}
\, \times \,
\alpha_1^{cl}! \, J_1^{-\alpha_1^{cl}}\, 
\times \, \Big(\sqrt{y(1-y)}\Big)^{-\alpha_1^{cl}}
\EN
where $\overline{k}_2=k_2-\alpha_1^{cl}$ and $\overline{k}_3
=k_3-\alpha_1^{cl}$ represent  
the number of impurities  in the operators 2 and 3, respectively, which contract with those of the operator 1. 
Note that the factor $\alpha_1^{cl}!$ comes from the 
permutations of the ordering of impurities contracting 
between 2 and 3. 

Secondly, each single contraction with momenta 
$p_i^{(1)}$ and $p_i^{(2)}$ respectively between 1 and 2 
after the summation over the permutation of their 
positioning gives the factor (suppressing the 
spacetime dependent factor $1/|\vec{x}_{12}|^2$)  
\EQ
\sum_{a=0}^{J_2}\exp\Big[2\pi i a ({p_i^{(2)}\over J_2} -
{p_i^{(1)}\over J_1})\Big]
\rightarrow 
J_1 
\e^{-\pi i y p_i^{(1)} }{\sin \pi y p_i^{(1)}\over \pi 
(p_i^{(1)}-{p_i^{(2)}\over y})}
\EN
in the large $J$ limit, 
which is expressed in terms of Neumann function as 
\EQ
J_1\sqrt{y} (-1)^{p_i^{(1)}+p_i^{(2)}+1}
\e^{-\pi i y p_i^{(1)} }\tilde{N}^{12}_{p_i^{(1)}p_i^{(2)}}.
\label{12cont}
\EN 
Here the Kronecker delta of the corresponding SO(4) indices is 
suppressed. 
Similarly, each single contraction with momenta 
$q_i^{(1)}$ and $q_i^{(3)}$ respectively between 1 and 3 
 gives
\EQ
\sum_{a=0}^{J_3}\exp\Big[2\pi i a (-{q_i^{(3)}\over J_3} +
{q_i^{(1)}\over J_1})\Big]
\rightarrow 
-J_1\sqrt{1-y} 
\e^{\pi i (1-y) q_i^{(1)} }\tilde{N}^{13}_{q_i^{(1)}q_i^{(3)}}.
\label{13cont}
\EN
On the other hand, each contraction with momenta 
$r_i^{(2)}$ and $r_i^{(3)}$ respectively between 2 and 3 
gives factor 1 (again apart from the spacetime factor) 
which is replaced by the corresponding Neumann functions using 
the identity
\EQ
1=(-1)^{r_i^{(2)}+1}\, 
4\pi \mu |\alpha_{(1)}|\sqrt{y(1-y)}\, \tilde{N}^{23}_{r_i^{(2)}r_i^{(3)}}\, . 
\label{23cont}
\EN

Collecting all these factors together, we conclude that 
the CFT coefficient is given as 
\[
C_{123}=( C^{I_1}C^{I_2}C^{I_3}) (-1)^{\alpha_1^{cl}+\alpha_2^{cl}
+\alpha_3^{cl}}
{\sqrt{J_1J_2J_3}\over N}
 \, \times \, J_1^{-k_1}\Big({J_2\over J_1}\Big)^{-\overline{k}_2/2}
\Big({J_3\over J_1}\Big)^{-\overline{k}_3/2}
\, \times \,
\alpha_1^{cl}! \, J_1^{-\alpha_1^{cl}}\, 
\]
\[
\times \, \Big(\sqrt{y(1-y)}\Big)^{-\alpha_1}
\Big(\prod_{12} 
J_1\sqrt{y}\tilde{N}^{12}_{p_i^{(1)}p_i^{(2)}}\Big)
\Big(\prod_{13} J_1\sqrt{1-y} \tilde{N}^{13}_{q_i^{(1)}q_i^{(3)}}\Big)
\]
\[
\times \Big(\prod_{23} 4\pi \mu |\alpha_{(1)}|\sqrt{y(1-y)}\, 
\tilde{N}^{23}_{r_i^{(2)}r_i^{(3)}}\Big)
\]
\[=( C^{I_1}C^{I_2}C^{I_3})  (-1)^{\alpha_1^{cl}+\alpha_2^{cl}+
\alpha_3^{cl}}
{\sqrt{J_1J_2J_3}\over N}\alpha_1! \Big({J_1\over 4\pi \mu |\alpha_{(1)}^{cl}|}
\Big)^{-\alpha_1^{cl}}
\]
\EQ
\times \Big(\prod_{12}  \tilde{N}^{12}_{p_i^{(1)}p_i^{(2)}}\Big)
\Big(\prod_{13}  \tilde{N}^{13}_{q_i^{(1)}q_i^{(3)}}\Big)
\Big(\prod_{23}
\tilde{N}^{23}_{r_i^{(2)}r_i^{(3)}}\Big)\, ,
\EN
where $\prod_{rs}$ denotes the product over the 
contractions and the symbol $( C^{I_1}C^{I_2}C^{I_3})$ 
denotes the part of  SO(4) contractions of polarization tensors 
associated with 
the free-field contractions .  
We note that the phase factors 
of the expressions (\ref{12cont}) $\sim$ (\ref{23cont}) 
cancel using the level-matching condition and that 
in terms of string notation, the following equality 
\[
( C^{I_1}C^{I_2}C^{I_3}) (-1)^{\alpha_1^{cl}+\alpha_2^{cl}+
\alpha_3^{cl}}\Big(\prod_{12}  \tilde{N}^{12}_{p_i^{(1)}p_i^{(2)}}\Big)
\Big(\prod_{13}  \tilde{N}^{13}_{q_i^{(1)}q_i^{(3)}}\Big)
\Big(\prod_{23}
\tilde{N}^{23}_{r_i^{(2)}r_i^{(3)}}\Big)\, 
\]
\EQ
=\, _{(1)}\langle C^{I_1}|\, _{(2)}\langle C^{I_2}|\, _{(3)}\langle C^{I_3}|
|E_a\rangle\, 
\label{stringrepofcont}
\EN
is valid.

Let us next confirm that this result matches 
 the string field theory through our holographic relation.  
 Since we have already 
expressed a main part of the CFT 
coefficient obtained on the gauge-theory side in terms of string states, 
only nontrivial part 
for this task is to examine the  prefactor. 
In the large $\mu$-limit, it takes the form
\[
P_{123}=\mu(\alpha_0^{(2)\dagger}\alpha_0^{(2)}+
\alpha_0^{(3)\dagger}\alpha_0^{(3)}
-\alpha_0^{(1)\dagger}\alpha_0^{(1)})
\] 
\[
+ {\mu\over 2} \sum_{m=1}^{\infty}
(\alpha_m^{(2)\dagger}\alpha_m^{(2)} + 
 \alpha_m^{(3)\dagger}\alpha_m^{(3)} 
- \alpha_m^{(1)\dagger}\alpha_m^{(1)}
+ [m\rightarrow -m] )
\]
\EQ
+{\mu\over 2} \sum_{m=1}^{\infty}
(\alpha_m^{(2)\dagger}\alpha_{-m}^{(2)} + 
 \alpha_m^{(3)\dagger}\alpha_{-m}^{(3)} 
- \alpha_m^{(1)\dagger}\alpha_{-m}^{(1)} 
+ [m\rightarrow -m] )
\EN
Obviously, the contributions from 
the first and second lines is simply determined by counting 
the difference 
of the numbers of impurities in the process 
$2+3\rightarrow 1$. Furthermore, in the third line, 
the impurity preserving part corresponding to 
contractions $1\leftrightarrow 2$ and $1\leftrightarrow 3$ cancels due to 
the equalities  $\tilde{N}_{-nm}^{rs}=\tilde{N}_{n-m}^{rs}, 
\tilde{N}_{nm}^{rs}=\tilde{N}_{-n-m}^{rs}$. 
Then, the property that $\tilde{N}^{23}_{mn}$ 
is independent of the signs of momenta $(m,n)$ 
shows that the contribution of the third term 
is $\mu \alpha_1^{cl}$. Thus all together the total contribution 
of the prefactor is $2\mu \alpha_1^{cl}$. 
But this is just canceled by the denominator factor 
$1/\mu (\Delta_2+ \Delta_3-
\Delta_1)$ of the holographic relation. 
Using (\ref{stringrepofcont}), 
it is now clear that the CFT coefficient precisely 
satisfies our holographic relation (\ref{clambdarelation1}) 
$\sim$ (\ref{clambdarelation3}) including the sign and 
numerical factors.

\section{Class I non-singlet vector impurities}
\setcounter{equation}{0}
In this section, we extend the result of the 
previous section to the situation where (non-singlet) 
vector excitations are involved. A vector excitation of momentum $p$  
at the position $a$ 
corresponds to the insertion of a 
derivative  $\e^{2\pi i a p/J} D_i$ $(i\in (1\sim 4))$ 
in the trace (\ref{genetracelessop}). We assume here that only 
traceless part are involved with respect to the vector 
SO(4) indices too. So, the vector SO(4) indices which will be mostly 
suppressed below should be understood to be contracted with traceless 
polarization tensors similarly as in the previous section.  As we have 
argued in detail in \cite{dy} on the basis of 
the GKP-Witten relation, the derivatives must then be 
computed by assuming the following form for the variation of 
the distance function $|\vec{x}_{rs}|$ under the shift of the 
spacetime coordinates $\vec{x}_r \rightarrow \vec{x}+\delta\vec{x}_r$, 
\EQ
{1\over |\vec{x}_{rs}|^{2}} \rightarrow 
{1 \over (\vec{x}_r-\vec{x}_s)^2
- 2\delta\vec{x}_r\cdot  \delta\vec{x}_s}
\label{bulkvariation}
\EN
in order to be consistent with the SO(4) symmetry and 
orthogonality of vector states. 
The invariants such as $(\delta\vec{x})^2$ can be ignored 
because of the traceless condition. 

This allows us to 
treat the pairs of vector indices almost as those 
of additional virtual 
scalar indices. Namely,  the pair of derivatives 
of the spacetime factor of a free propagator can 
contribute in the form 
\[
\sum_s \e^{2\pi i a ({p^{(r)}\over J_r}\pm {q^{(s)}\over J_s})}
\partial_i^{(r)}\partial_j^{(s)}{1\over |\vec{x}_{rs}|^2}
\sim {2\over |\vec{x}_{rs}|^4}\delta_{ij}\tilde{N}^{rs}_{p^{(r)},\pm q^{(s)}}
\times \, \, \mbox{phase factor}
\]
where the phase factor depending on  $(r,s)$ is the same as for the case of scalars. 
Thus a pair of vector derivatives yields essentially the same factor 
as in the case of contractions of scalar impurities, 
except for the factor 2 which is absorbed in the normalization,  
provided that they act on the free propagator. 
This implies that the derivatives acting on $1\leftrightarrow 2$ and 
$1\leftrightarrow 3$ contractions give the same factors on both 
sides of gauge theory and string field theory. 
In other words, the holographic relation is satisfied 
as it stands when vector impurities are conserved for 
arbitrary impurity non-preserving processes in which impurity non-preserving 
contractions between 2 and 3 only occur for 
scalar excitations.  

Therefore, in the rest of this section, it is sufficient to 
concentrate to the case where all vector impurities contribute to 
$(2\leftrightarrow 3)$ contractions. Suppose first that we add $\beta_1^{cl}$ 
such derivatives for a process with nonzero 
scalar exchange in the $2-3$ channel ($\alpha_1^{cl}>0$). 
Since the spacetime factor 
${1/|\vec{x}_{23}|^{2\alpha_1^{cl}}}$ only comes from 
a single group of adjacent scalar impurities, 
the derivatives act as
\[
{1\over  (2J_2)^{\beta_1^{cl}/2}
(2J_3)^{\beta_1^{cl}/2}}\times  {1\over (\beta_1^{cl})!}(\partial^{(2)})^{\beta_1^{cl}}
(\partial^{(3)})^{\beta_1^{cl}}
{1\over |\vec{x}_{23}|^{2\alpha_1^{cl}}}
\]
where the first factor comes from the 
normalization constant associated with the vector 
excitations. 
 For notational 
simplicity, we suppress the SO(4) vector indices. 
The denominator factor $1/(\beta_1^{cl})!$ is 
the symmetry factor to cancel the over-counting. This is equal to 
\[
J_1^{-\beta_1^{cl}}
\Big({J_2\over J_1}\Big)^{-\beta_1^{cl}/2}
\Big({J_3\over J_1}\Big)^{-\beta_1^{cl}/2}
\times 
{\Gamma(\alpha_1^{cl}+\beta_1^{cl})\over 
\Gamma(\alpha_1^{cl})}
{1\over |\vec{x}_{23}|^{2(\alpha_1^{cl}+\beta_1^{cl})}}
\]
\EQ
=J_1^{-\beta_1^{cl}}
\Big(\sqrt{y(1-y)}\Big)^{-\beta_1^{cl}}
{\Gamma(\alpha_1^{cl}+\beta_1^{cl}+1)\over 
\Gamma(\alpha_1^{cl}+1)}{2\alpha_1^{cl}\over 
2(\alpha_1^{cl}+ \beta_1^{cl})}
{1\over |\vec{x}_{23}|^{2(\alpha_1^{cl}+\beta_1^{cl})}} \, .
\EN

As in the scalar case, we further multiply the factor 
$$1=(-1)^{r_i^{(2)}+1}4\pi \mu |\alpha_{(1)}|\sqrt{y(1-y)}\, 
\tilde{N}^{23}_{r_i^{(2)}r_i^{(3)}}$$ corresponding to each pair 
of derivatives $\partial^{(2)}\partial^{(3)}$. By this procedure, the 
power of $y(1-y)$ is canceled and replaced by the 
Neumann functions for $(2-3)$ channel, 
leaving us the correction factor 
$\Big({J_1\over 4\pi \mu |\alpha_{(1)}^{cl}|}
\Big)^{-\beta_1^{cl}}$. Furthermore, the factor 
${\Gamma(\alpha_1^{cl}+\beta_1^{cl}+1)\over 
\Gamma(\alpha_1^{cl}+1)}{2\alpha_1^{cl}\over 
2(\alpha_1^{cl}+ \beta_1^{cl})}$ is just 
the one required for modifying the 
energy factor in the holographic relation as 
\[
{\Gamma(\alpha_1^{cl}+1)\over 2\alpha_1^{cl}} 
\rightarrow {\Gamma(\alpha_1^{cl}+\beta_1^{cl}+1)
\over 2(\alpha_1^{cl}+\beta_1^{cl})} 
\]
corresponding to the shift of $\Delta_2+\Delta_3-\Delta_1$ 
due to the addition of vector impurities in the $(2-3)$ channel. 

Turning now to the string side, only difference for 
vector excitations from the purely scalar case 
is that the contribution of vector modes to the prefactor consists only of 
$\sin$ modes, 
\[
{\mu\over 2} \sum_{m=1}^{\infty}
(\alpha_m^{(2)\dagger}\alpha_m^{(2)} + 
 \alpha_m^{(3)\dagger}\alpha_m^{(3)} 
- \alpha_m^{(1)\dagger}\alpha_m^{(1)}
+ [m\rightarrow -m] )
\]
\EQ
-{\mu\over 2} \sum_{m=1}^{\infty}
(\alpha_m^{(2)\dagger}\alpha_{-m}^{(2)} + 
 \alpha_m^{(3)\dagger}\alpha_{-m}^{(3)} 
- \alpha_m^{(1)\dagger}\alpha_{-m}^{(1)} 
+ [m\rightarrow -m] ).
\label{vecprefactor}
\EN
Then, the independence of $(2-3)$ Neumann functions on the momentum 
(except for the sign factor) shows that the vector contribution 
in this channel is zero due to cancellation between the first 
and the second line in this expression. Hence, it does not 
contribute at all to the modification of the 3-point function. 
Also it is easy to check that the sign and phase factors exactly 
match using level matching condition, 
in the same way as purely scalar case of the 
previous section. 
Thus we can conclude that the holographic relation is precisely satisfied. 

The case with $\alpha_1^{cl}=0$ requires a separate consideration. 
In this case, there is no direct contraction  
for scalar fields in the $(2-3)$ channel. Also 
the correction factor before taking derivatives 
is singular in the large $\mu$-limit, and 
we have to take into account mixing with double-trace operators.  
The space-time dependence 
come from the order $\lambda'$ contributions to the anomalous dimensions. 
Let us proceed inductively, starting from $\beta_1^{cl}=1$. 
We have to consider 
\[
{1\over (2J_2)^{1/2}
(2J_3)^{1/2}}\times \partial^{(2)}\partial^{(3)}
{1\over |\vec{x}_{23}|^{2\alpha_1}}\sim 
(J_2J_3)^{-1/2}\alpha_1{1\over |\vec{x}_{23}|^2}
\]
with $\alpha_1 \sim O((\mu \alpha_{(1)})^{-2})$. 
The factor $\alpha_1$ is precisely the necessary factor 
which modifies the singular energy factor into 
non-singular one corresponding to nonzero $\beta^{cl}_1$, 
\[
{\Gamma(\alpha_1+1)\over 2\alpha_1} \sim 
{\Gamma(\alpha_1^{cl}+1)\over 2\alpha_1}
\rightarrow \alpha_1 \times {\Gamma(\alpha_1^{cl}+1)\over 2\alpha_1}
=
{\Gamma(\alpha_1^{cl}+\beta_1^{cl}+1)
\over 2(\alpha_1^{cl}+\beta_1^{cl})} ={1\over 2} \, , 
\]
with $\beta_1^{cl}=1$ and $\alpha_1^{cl}=0$, and the 
power $(J_2J_3)^{-1/2}$ plays the same role generating the 
$(2-3)$ Neumann functions as in the 
above case with 
nonzero $\alpha_1^{cl}$. Since there does not occur any correction factor 
from the vector prefactor on the string side, this proves the validity of 
the holographic relation. Note that in this special case the mixing with 
double-trace operator for the operator 1 is effective on the gauge-theory 
side, but on the string side the $(2-3)$ Neumann functions 
contribute the same common factor for both $|H_3\rangle_D$ 
and $|H_3\rangle_{SV}$. 
Once the spacetime factor $1/|\vec{x}_{23}|^2$ is 
generated correspondingly to a single vector exchange as above, 
we can apply the same argument inductively at each time 
of adding one vector exchange between 2 and 3 as we have 
discussed for the case starting from nonzero $\alpha_1^{cl}$.  
This completes our argument for the holographic relation 
for non-singlet vector impurities.

\section{Singlet operators and class II processes}
\setcounter{equation}{0}
All previous arguments are restricted to cases where 
there is no trace part in the external lines. When 
we allow trace part for the 
SO(4)  representation, the situation becomes more 
cumbersome and it is not easy to make treatments in a 
general way because of various possible mixings among different 
configurations of operator products 
even at the level of single trace operators. 
It is also important that the inclusion of singlet 
opens the possibility of class II processes. 
In the present paper, we study several typical cases 
in order to convince ourselves that our holographic 
relation must be valid in such cases too. In particular, 
we will see that various characteristic structures of the 
CFT coefficients  
depending differently on scalar and vector excitations are 
nicely captured by the interplay between the factor $G$ in the 
holographic relation and 
the very specific prefactor of our string-interaction vertex. 

\subsection{Class I with scalar singlets}
The simplest (normalized) BMN operator with singlet representation 
for scalar excitation \cite{PR} 
is 
\EQ
O_{ss (p; -p)}={1\over 2\sqrt{J N^{J+2}}}
\Tr
\Big[\sum_{a=0}^{J}
\phi_i Z^a \phi_iZ^{J-a}\, \e^{-{2\pi i\over J}ap}
-4 \overline{Z} Z^{J+1}
\Big] 
\EN
where the scalar index $i$ is summed over from 
$i=5$ to $8$ and $p\ne 0$.  
When $p=0$, the normalization constant should be 
replaced by ${1\over 2\sqrt{2J N^{J+2}}}$. 
The necessity of the mixing of the second term can be 
easily understood when $p=0$:  In order for this operator 
to be BPS, it must be symmetric and traceless with respect to 
SO(6) symmetry. The role of the second term is to meet this requirement by completing the vanishing SO(6) trace from the $Z$-$\overline{Z}$ directions. 

Let us start from the case where the operator 1 and 
3 are in the ground state, 
\EQ
O^{(1)}={1\over \sqrt{J_1 N^{J_1}}}\Tr
[Z^{J_1}], \quad 
O^{(3)}={1\over \sqrt{J_3 N^{J_3}}}\Tr
[Z^{J_3}] \, , 
\label{ground13}
\EN
and the operator 2 is the above singlet-state operator, 
\EQ
O^{(2)}_{s (p; -p)}={1\over 2\sqrt{J_2 N^{J_2+2}}}
\Tr
\Big[\sum_{a=0}^{J_2}
\phi_i Z^a \phi_iZ^{J_2-a}\, \e^{-{2\pi i\over J_2}ap}
-4 \overline{Z}Z^{J_2+1}
\Big] \, .  
\label{singlet2}
\EN 
On the gauge-theory side,  the first term 
of (\ref{singlet2}) has no contribution  in the leading 
order in the perturbation theory. The second term 
can contribute in the leading order if and only if one of 
the $\overline{Z}$ in $O^{(2)}_{s (p; -p)}$ 
is contracted with one of $Z$ in $O^{(3)}$ as in fig. 2. The mixing with 
double-trace operator can again be ignored. 
Thus, apart from the spacetime factor (which takes 
the correct form), 
the CFT coefficient is given as 
\EQ
C_{123}=-4 \times {1\over 2\sqrt{J_2N^{J_2+2}}}
\times {1\over \sqrt{J_1N^{J_1}J_3N^{J_3}}}J_3 J_1 
N^{J_1}=
-2{\sqrt{J_1J_2J_3}\over N} {1\over J_2}\, . 
\EN

On the string side, the state corresponding to 
the operator 2 is
$$\, _{(2)}\langle 0| {1\over 2}
\alpha^{(2)}_{p, i}\alpha^{(2)}_{-p, i}\, .$$
Using the explicit form of 
the $2-2$ Neumann function $\tilde{N}_{mn}^{22}=
{(-1)^{m+n}\over 4\pi \mu|\alpha_{(1)}|y}$, 
we find that the prefactor is simply 
$2\mu \alpha_1^{cl}=2\mu$ and hence the 
matrix element of the interaction vertex is given as 
\EQ
4\mu{\sqrt{J_1J_2J_3}\over N}(-\tilde{N}^{22}_{p,-p}) 
=-4\mu {\sqrt{J_1J_2J_3}\over N}{1\over 4\pi \mu|\alpha_{(1)}|y}\, .
\EN
By multiplying the factor $G$
\EQ
{1\over \mu(\Delta_2+\Delta_3-\Delta_1)} 
\Big(f{J_2J_3 \over J_1}\Big)^{-1}
\Gamma({\Delta_2+\Delta_3-\Delta_1 \over 2}+1)
\sim {1\over 2\mu}{4\pi \mu |\alpha_{(1)}|\over J_1}, 
\EN
we find that the holographic relation is precisely satisfied 
including numerical coefficient and sign.  
\begin{center}
\begin{figure}
\begin{picture}(100,100)
\put(170,0){
\includegraphics[width=100pt]{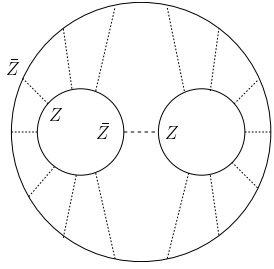}
}
 \end{picture}
\caption{The contraction of $Z$ and $\overline{Z}$ 
between 2 and 3, with 2 being the singlet operator and  
3 the ground state operator. }
\label{fig2} 
\end{figure}
\end{center}

\vspace{-1.3cm}
This simplest example  shows that the mixing 
of the second term (\ref{singlet2}) is absolutely 
necessary for this agreement, and also that 
it is responsible to the 
  (2-2) Neumann functions  on the 
string side. 
To convince the universality of this role of the 
$Z\overline{Z}$ mixing term, it is useful to treat the 
case where both of the operators 2 and 3 are singlet states. 
So, now $O^{(3)}$ in (\ref{ground13}) is replaced by
\EQ
O^{(3)}_{s (q; -q)}={1\over 2\sqrt{J_3 N^{J_3+2}}}
\Tr
\Big[\sum_{a=0}^{J_3}
\phi_i Z^a \phi_iZ^{J_3-a}\, \e^{-{2\pi i\over J_3}aq}
-4 \overline{Z}Z^{J_3+1}
\Big] \, .  
\EN 
Then, the CFT coefficient consists of two 
contributions with different sets of free-field 
contractions as in fig. 3, 
\EQ
C_{123}=C_{123}^{ii\leftrightarrow jj} +
C_{123}^{Z\overline{Z}}
\label{cdecomp}
\EN
corresponding  to the contributions from the contractions 
between scalar excitations and between $Z, \overline{Z}$ fields,  respectively. By similar calculations as above, we find
\EQ
C_{123}^{ii\leftrightarrow jj}={\sqrt{J_1J_2 J_3}\over N}
{4 \over J_2J_3}, 
\quad 
C_{123}^{Z\overline{Z}}
={\sqrt{J_1J_2 J_3}\over N}
{8 \over J_2J_3}. 
\EN 

\begin{center}
\begin{figure}
\begin{picture}(100,100)
\put(90,0){
\includegraphics[width=100pt]{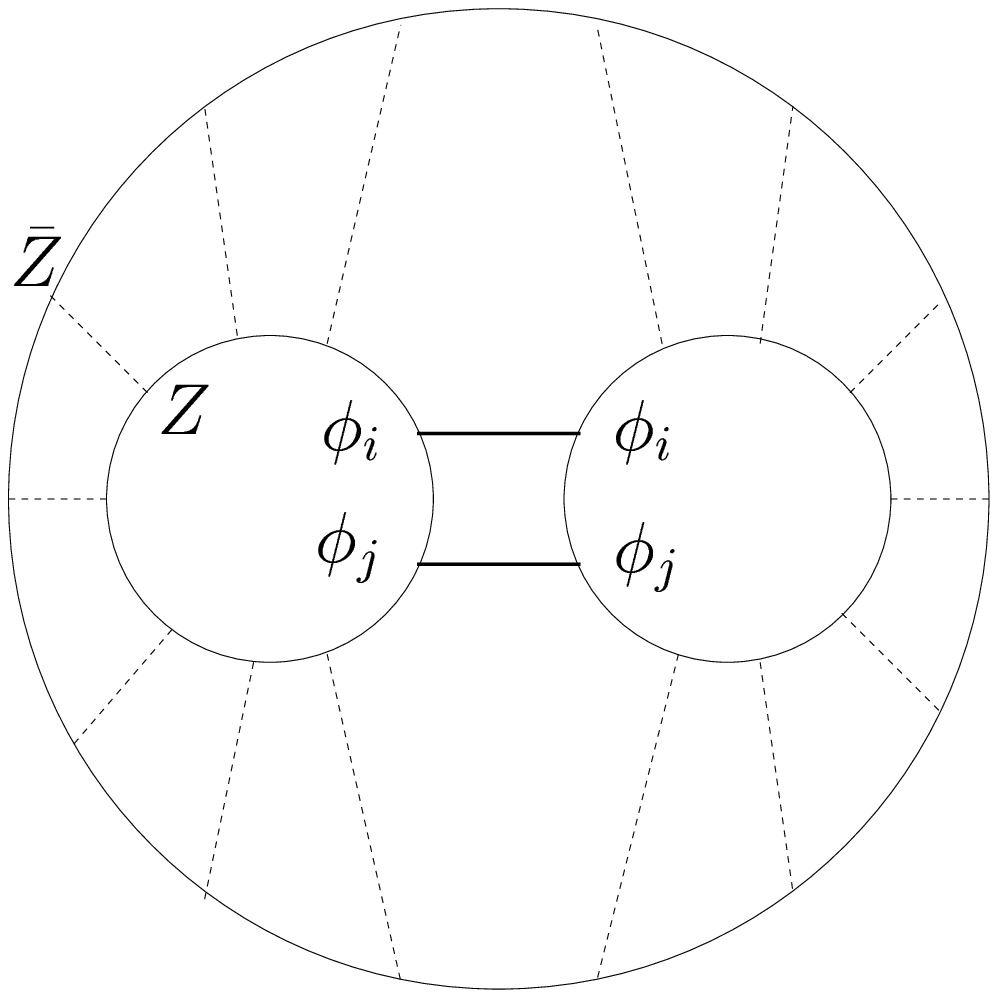}
}
\put(250,0){
\includegraphics[width=100pt]{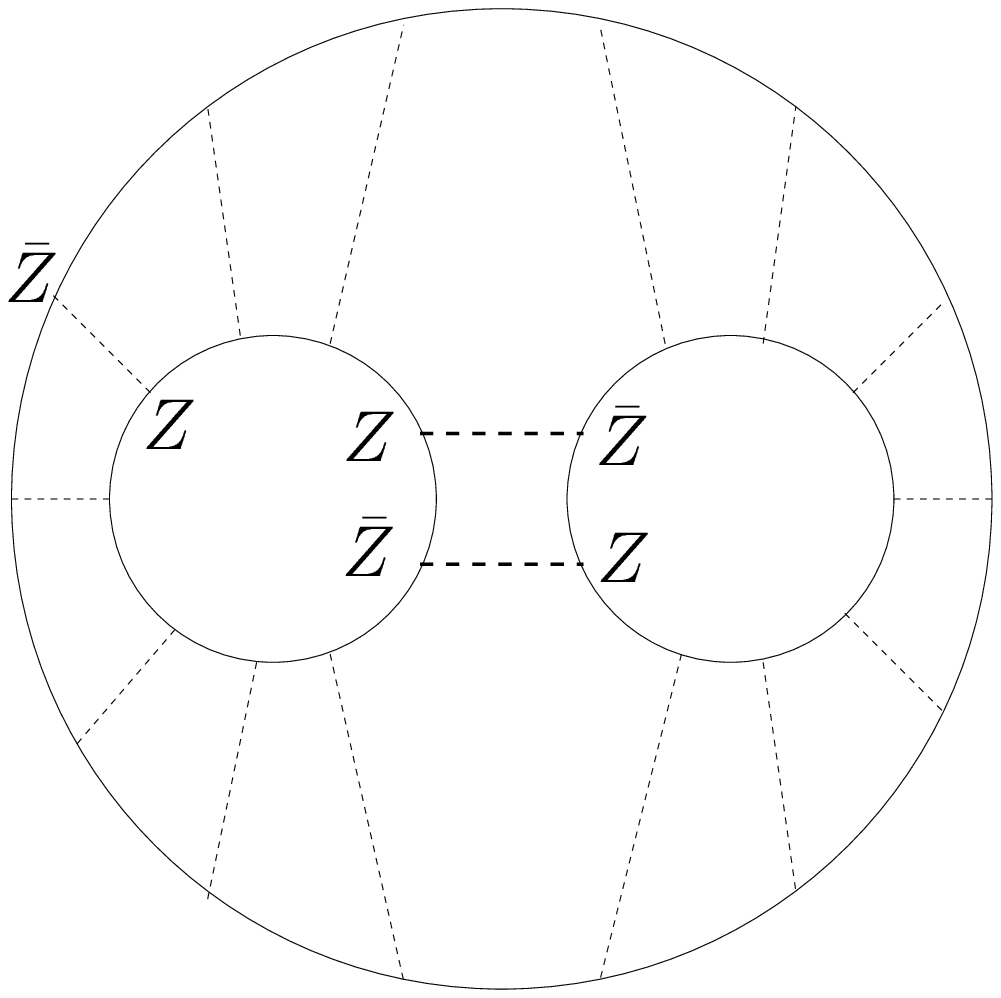}
}
\end{picture}
\caption{The two types  of contractions 
between 2 and 3 which are both singlet operators with 
two scalar impurities.  }
\label{fig3} 
\end{figure}
\end{center}

On the string-field side, we find the following 
matrix element of the interaction vertex, by using self-explanatory 
notations,  
\EQ
\lambda_{123}=\lambda_{123}^{2\leftrightarrow 3}
+\lambda_{123}^{self}
\label{lambdadecomp}
\EN
with
\EQA
\lambda_{123}^{2\leftrightarrow 3}
&&\hspace{-0.5cm}=4\mu{\sqrt{J_1J_2J_3}\over N}\times 2
(\tilde{N}^{23}_{pq})^2=4\mu{\sqrt{J_1J_2J_3}\over N}\times 2\Big({J_1\over 4\pi \mu |\alpha_{(1)}|}\Big)^2
{1\over J_2J_3},
\\
\lambda_{123}^{self}
&&\hspace{-0.5cm}=4\mu{\sqrt{J_1J_2J_3}\over N}
\times 4\tilde{N}^{22}_{p,-p}\tilde{N}^{33}_{q,-q}
=4\mu{\sqrt{J_1J_2J_3}\over N}
\times 4\Big({J_1\over 4\pi \mu |\alpha_{(1)}|}\Big)^2
{1\over J_2J_3}\, 
\EQN
where we have again 
used the property that the Neumann functions 
appearing here are independent of the signs of momentum. 
Since the  factor $G$ is 
\[
{1\over \mu(\Delta_2+\Delta_3-\Delta_1)} 
\Big(f{J_2J_3 \over J_1}\Big)^{-2}
\Gamma({\Delta_2+\Delta_3-\Delta_1 \over 2}+1)
\sim
2{1\over 4\mu}
\Big({4\pi \mu |\alpha_{(1)}|\over J_1}\Big)^2\, , 
\]
the results on both sides exactly match separately 
for two contributions between (\ref{cdecomp}) and 
(\ref{lambdadecomp}). These two exercises seem sufficient 
to convince the general validity of our holographic relation 
for class I processes involving singlet representation. 

\subsection{Class II with scalar singlets}
Let us next consider  an example of  class II processes which 
are possible with the presence of trace part 
in the operator 1. Consider the simplest nontrivial 
case with 
\EQ
O^{(1)}_{s (p; -p)}={1\over 2\sqrt{J_1N^{J_1+2}}}
\Tr
\Big[\sum_{a=0}^{J_1}
\phi_i Z^a \phi_iZ^{J_1-a}\, \e^{-{2\pi i\over J_1}ap}
-4 \overline{Z}Z^{J_1+1}
\Big] \, 
\label{singlet1}
\EN
\EQ
O^{(2)}={1\over \sqrt{J_2 N^{J_2}}}\Tr
[Z^{J_2}]\, , \quad 
O^{(3)}={1\over \sqrt{J_3 N^{J_3}}}\Tr
[Z^{J_3}] \, .
\EN
On the gauge-theory side, there is no possibility 
of free contraction for either term of the operator 1. 
This means that the CFT coefficient vanish in the 
leading order at least in the $1/\mu$-expansion and hence is at most 
of order $g_2\lambda' \propto 1/\mu^2$, if it does not vanish. 

On the other hand, the corresponding 
matrix element of the 3-point interaction vertex 
is given as, suppressing an obvious SO(4) indices, 
\[
{\sqrt{J_1J_2J_3}\over N}
{\mu \over 2}\langle 0|\, \alpha_p^{(1)}\alpha_{-p}^{(1)}
(2+\alpha_p^{(1)\dagger}\alpha_{-p}^{(1)}+
\alpha_{-p}^{(1)\dagger}\alpha_p^{(1)}) 
\]
\[
\times \exp\Big[
-{1\over 2}\Big(\tilde{N}^{11}_{pp}
(\alpha_p^{(1)\dagger})^2 +\tilde{N}^{11}_{-p,-p}(
\alpha_{-p}^{(1)\dagger})^2
\Big)
-\tilde{N}^{11}_{p, -p}\alpha_p^{(1)\dagger} 
\alpha_{-p}^{(1)\dagger}
\Big]\, |0\rangle
\]
\EQ
=-{\sqrt{J_1J_2J_3}\over N}
{\mu \over 2}\, (2\tilde{N}^{11}_{p, -p} +\tilde{N}^{11}_{-p,-p}
+\tilde{N}^{11}_{pp})\, .
\EN
The explicit form of the $(11)$-Neumann function is 
\EQ
\tilde{N}^{11}_{nm}=\tilde{N}_{-n, -m}^{11}
=-\tilde{N}^{11}_{-n, m}=-\tilde{N}^{11}_{n, -m}
=(-1)^{m+n+1}
{\sin (\pi my) \sin (\pi n y)\over \pi \mu |\alpha_{(1)}|}
\EN
in the leading order in the $1/\mu$ expansion. 
The correction terms for this expression 
is at most of order $1/(\mu |\alpha_{(1)}|)^3$. 
Therefore, the matrix element vanishes in the leading 
order and possible corrections must be at most 
of order 
${\mu \over (\mu |\alpha_{(1)}|)^3}$. 

Let us next examine the multiplying factor, 
\[
G\equiv {1\over \mu(\Delta_2+\Delta_3-\Delta_1)} 
\Big(f{J_2J_3 \over J_1}\Big)^{-{\Delta_2+\Delta_3-\Delta_1 \over 2}}
\Gamma({\Delta_2+\Delta_3-\Delta_1 \over 2}+1)\, . 
\]
One of new features of class II processes is that $
\alpha_1^{cl} ={\Delta_2^{cl}+\Delta_3^{cl}-\Delta_1^{cl} \over 2}$ is a negative integer. In the present example, 
\[
\alpha_1^{cl}=-1
\]
which leads to the following leading behavior 
\EQ
G\sim {1\over \mu}
f{J_2J_3\over J_1}{(\mu\alpha_{(1)})^2 \over p^2}
= {1\over \mu}
{J_1\over 4\pi\mu|\alpha_{(1)}|}{(\mu\alpha_{(1)})^2 \over p^2}=
{J_1 |\alpha_{(1)}| \over 4\pi p^2} \, . 
\label{correctionfactorg}
\EN
Note that we have here taken into account the 
singularity of the $\Gamma$-function $\Gamma(\epsilon) 
\sim 1/\epsilon$. 
Thus we conclude that the CFT coefficient must be 
at most of order $1/(\mu\alpha_{(1)})^2 \sim \lambda'$ 
in conformity with the result from the gauge-theory side. 
We have again seen the crucial role played by the 
 factor $G$ in relating both sides. 
The precise calculation of the higher order terms 
for 3-point correlation functions 
would require in general to take into account 
 various mixing terms involving 
both bosonic and fermionic impurities.  

\subsection{Vector singlets}

The case of vector-singlet states is more subtle. 
The general prescription adopted in section 4 for dealing with traceless 
vector excitations is not 
applicable to this case, since in the presence of 
trace part there are a variety of 
different ways of constructing SO(4) invariants after 
taking spacetime derivatives 
comparing the case for purely non-singlet processes. 
We consider only 
some simple examples. 

The simplest singlet operator with vector impurities 
with nonzero momenta $p$ and $-p$ is 
\EQ
O_{vs(p; -p)}={1\over 4\sqrt{JN^J}}\sum_{a=0}^{J-2}
\Tr\Big[
(D_{j}Z) Z^a (D_{j}Z) Z^{J-2 -a}\e^{-{2\pi i \over J-1}ap}
\Big]
\EN 
where the vector SO(4) index $j$ is summed over from 
1 to 4. The derivatives should now be computed 
in the standard way. It is easy to check that this satisfies 
the orthonormality.  Note that we can 
ignore the second derivatives $D^2Z$ in the present 
approximation due to the equation of motion. 

Let us  consider the simplest example of class I, 
in which the operator 2 is this singlet state, while 
1 and 3 are in the ground state, 
\EQ
O^{(2)}_{vs(p; -p)}={1\over 4\sqrt{J_2N^J_2}}\sum_{a=0}^{J_2-2}
\Tr\Big[
(D_{j}Z) Z^a (D_{j}Z) Z^{J_2-2 -a}\e^{-{2\pi i \over J_2-1}ap}
\Big]
 \label{vsinglet2}.
\EN
Then, the free-field contractions between 
1 and 2 yields the sum 
\[
\sum_{a=0}^{J_2-2}\e^{-{2\pi i \over J_2-1}ap}=0 \, .
\quad (p\ne 0) 
\]
We conclude that the corresponding 
CFT coefficient must be zero in the leading order. 
On the side of string-field theory, 
the matrix element of the 3-point vertex is 
\EQA
-{\sqrt{J_1J_2J_3}\over N}{\mu\over 2}\langle 0| \alpha_{p}^{(2)}\alpha_{ -p}^{(2)} 
&&\hspace{-0.6cm}
(2-\alpha_p^{(2)\dagger}\alpha_{-p}^{(2)}
-\alpha_{-p}^{(2)\dagger}\alpha_p^{(2)}) \nonumber \\
&&\hspace{-3cm}\times\exp \Big[
-{1\over 2}\tilde{N}^{22}_{pp}(\alpha_p^{(2)\dagger})^2 
-{1\over 2}\tilde{N}^{22}_{-p,-p}(\alpha_{-p}^{(2)\dagger})^2
-\tilde{N}^{22}_{p,-p}\alpha_p^{(2)\dagger}
\alpha_{-p}^{(2)\dagger}
\Big]|0\rangle \, 
\label{vsmatrixele1}
\EQN
which is equal to 
\EQ
-{\sqrt{J_1J_2J_3}\over N}2\mu(-2 \tilde{N}^{22}_{p, -p} + \tilde{N}^{22}_{pp}
+\tilde{N}^{22}_{-p, -p})=0
\EN
by using explicit forms of the Neumann functions which has 
already been used for the case of scalar singlet. 
Note that the crucial 
relative sign in the prefactor in the first line in the 
expression (\ref{vsmatrixele1}), which is opposite to 
the corresponding scalar case. Since there is no 
singularity in the factor $G$ in the holographic 
relation, this shows that the both sides match. 
It is easy to check that the vanishing of the CFT 
coefficient continues to be valid when we increase the 
number of singlet vector impurities 
in the lines 2 and 3 on both sides. 

The situation drastically changes for the 
class II case. Consider the simplest such process where 
the operator 1 is the vector singlet state with 2 and 3 being in the ground state:
\EQ
O^{(1)}_{vs(p; -p)}={1\over 4\sqrt{J_1N^{J_1}}}\sum_{a=0}^{J_1-2}
\Tr\Big[
(D_{j}Z) Z^a (D_{j} \overline{Z}) Z^{J_1-2 -a}\e^{-{2\pi i \over J_1-1}ap}
\Big]\, .
 \label{vsinglet1}
\EN
On the gauge-theory side, the 3-point function 
consists of three different types of contributions
\EQ
\langle \overline{O}^{(1)}_{vs(p; -p)}(\vec{x}_1)
O^{(2)}(\vec{x}_2)O^{(3)}(\vec{x}_3)\rangle =F_{123}^{1\leftrightarrow 2} 
+F_{123}^{1\leftrightarrow 3}
+F_{123}^{1\leftrightarrow (2,3)}
\EN
\begin{center}
\begin{figure}
\begin{picture}(100,80)
\put(40,0){
\includegraphics[width=110pt]{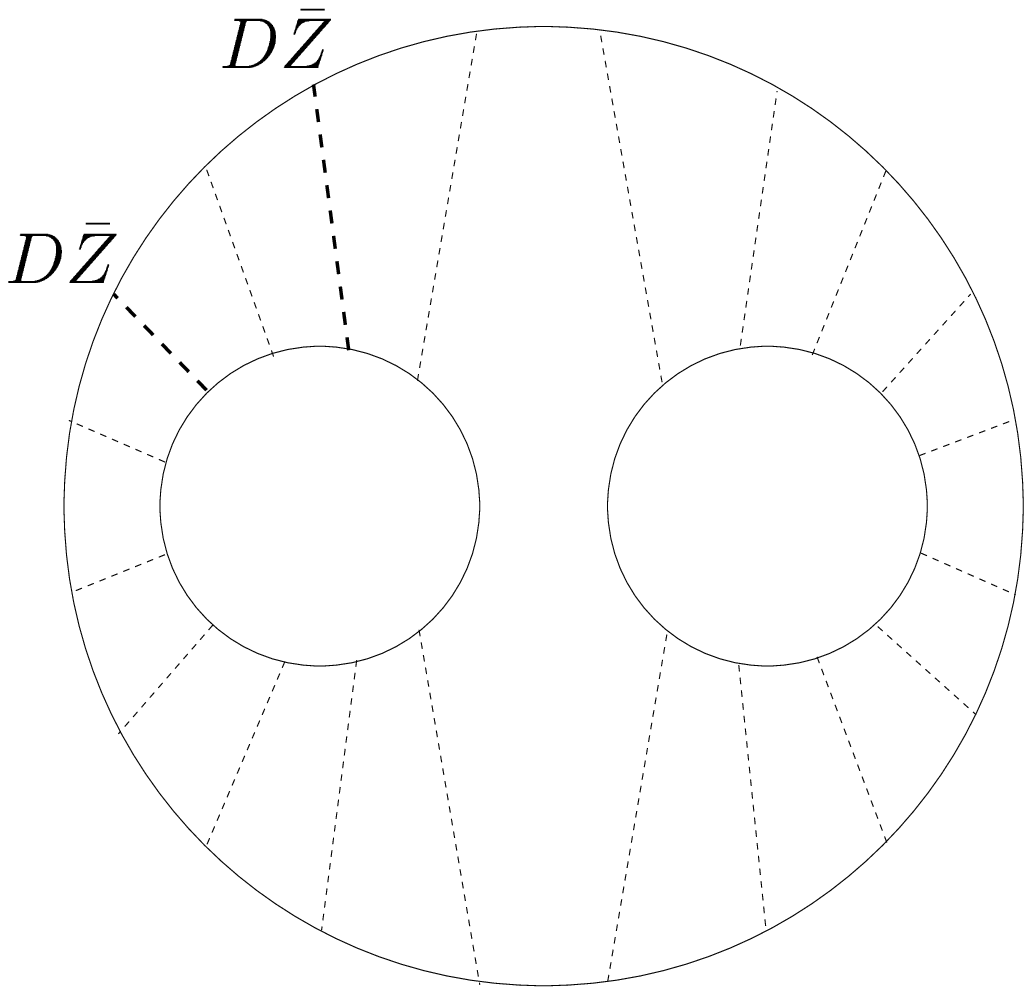}
}
\put(180,0){
\includegraphics[width=110pt]{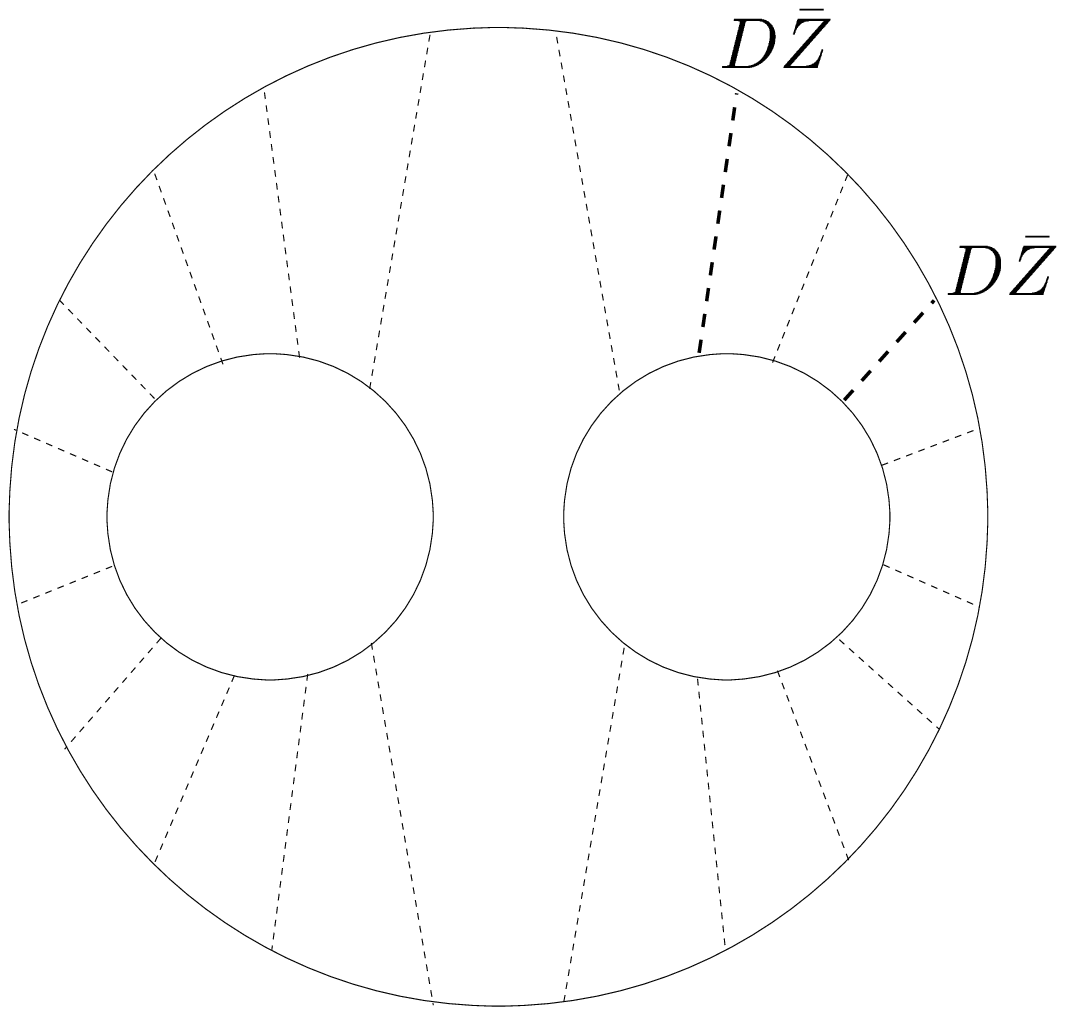}
}
\put(310,0){
\includegraphics[width=110pt]{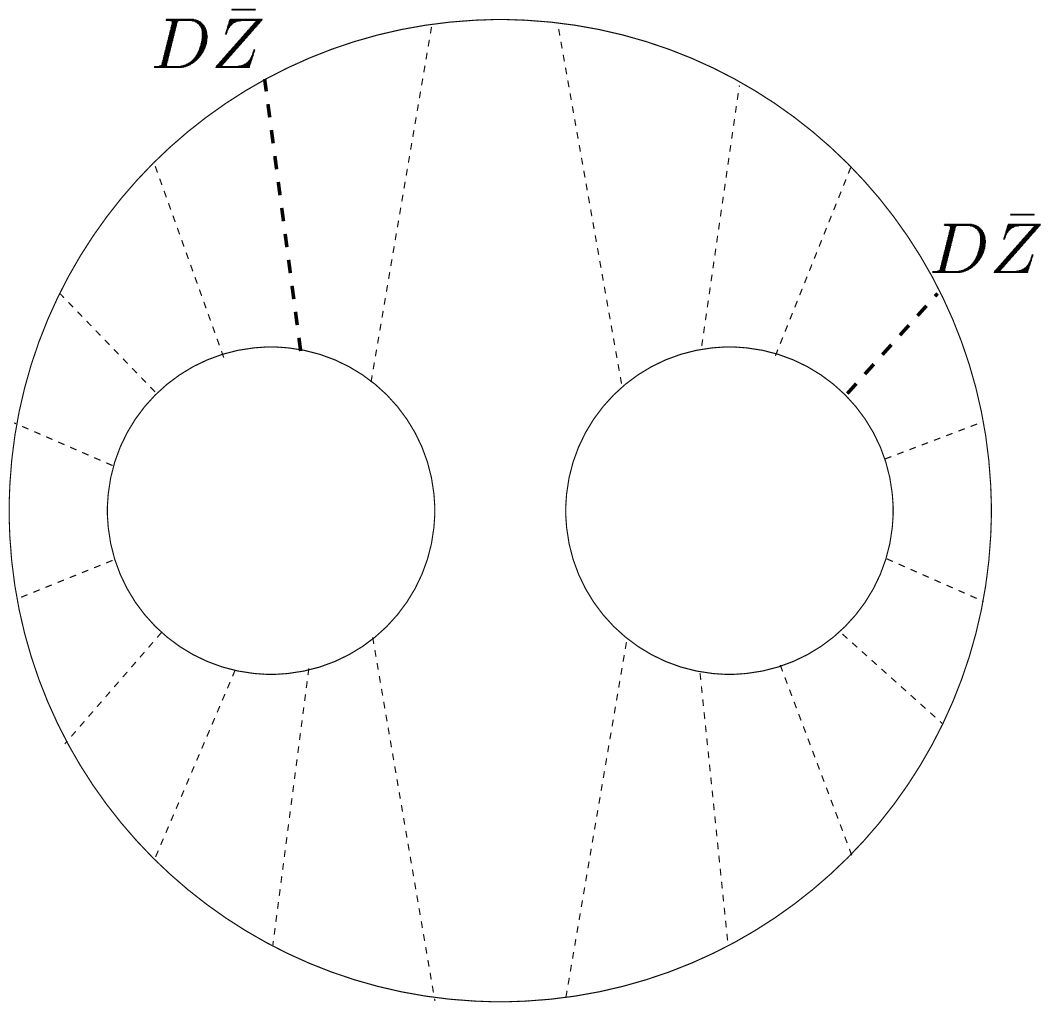}
}
 \end{picture}
\caption{The three types of contractions for 
the class II process with singlet vector operator 
for the line 1. }
\label{fig4} 
\end{figure}
\end{center}
In the first and second terms, both of two derivatives 
in (\ref{vsinglet1}) acts only on 
$(1\leftrightarrow 2)$ contractions and 
on $(1\leftrightarrow 3)$ contractions, respectively. 
In the third contribution, one of the 
derivatives acts on $(1\leftrightarrow 2)$ contractions 
and the other on $(1\leftrightarrow 3)$ contractions. See fig. 4. 
We find 
\EQA
F_{123}^{1\leftrightarrow 2} 
&&\hspace{-0.3cm}= {2\over N\sqrt{J_1J_2J_3}}J_2J_3 \times J_1^2
{\sin^2\pi yp\over (\pi p)^2}
{1\over |\vec{x}_{12}|^{2J_2+2}|\vec{x}_{13}|^{2J_3}}\, , \\
F_{123}^{1\leftrightarrow 3}
&&\hspace{-0.3cm}={2\over N\sqrt{J_1J_2J_3}}J_2J_3 \times J_1^2
{\sin^2\pi (1-y)p\over (\pi p)^2}
{1\over |\vec{x}_{12}|^{2J_2}|\vec{x}_{13}|^{2J_3+2}}\, , 
\EQN
\[
\hspace{-6.3cm}F_{123}^{1\leftrightarrow (2,3)}
={4\over N\sqrt{J_1J_2J_3}}J_2J_3 \]
\EQ
\hspace*{5cm}\times J_1^2
(-1)^p {\sin{\pi yp} \sin\pi (1-y)p\over (\pi p)^2}
{(\vec{x}_{12}\cdot \vec{x}_{13}) \over 
|\vec{x}_{12}|^{2J_2+2}|\vec{x}_{13}|^{2J_3+2}}\, .
\EN
Summing up these three contributions, we have 
the expression 
\[
2{\sqrt{J_1J_2J_3}\over N}J_1 {\sin^2\pi yp\over (\pi p)^2}
{|\vec{x}_{23}|^2 \over |\vec{x}_{12}|^{2J_2+2}|\vec{x}_{13}|^{2J_3 +2}}\, .
\]
The spacetime factor is precisely the form we should 
expect for a type II process with $\alpha_1^{cl}=-1$, and 
hence the CFT coefficient is 
\EQ
C_{123}=2{\sqrt{J_1J_2J_3}\over N}J_1 {\sin^2\pi yp\over (\pi p)^2}. 
\label{vscft1}
\EN

Let us turn to the string side. The matrix element 
of the interaction vertex is given by
\[
-{\sqrt{J_1J_2J_3}\over N}
{\mu \over 2}\langle 0|\, \alpha_p^{(1)}\alpha_{-p}^{(1)}
(2-\alpha_p^{(1)\dagger}\alpha_{-p}^{(1)}-
\alpha_{-p}^{(1)\dagger}\alpha_p^{(1)}) 
\]
\[
\times \exp\Big[
-{1\over 2}\Big(\tilde{N}^{11}_{pp}
(\alpha_p^{(1)\dagger})^2 +\tilde{N}^{11}_{-p,-p}(
\alpha_{-p}^{(1)\dagger})^2
\Big)
-\tilde{N}^{11}_{p, -p}\alpha_p^{(1)\dagger} 
\alpha_{-p}^{(1)\dagger}
\Big]\, |0\rangle
\]
\EQ
=-4{\sqrt{J_1J_2J_3}\over N}
{\mu \over 2}\, (-2\tilde{N}^{11}_{p, -p} +\tilde{N}^{11}_{-p,-p}
+\tilde{N}^{11}_{pp})={\sqrt{J_1J_2J_3}\over N}{8 \over \pi|\alpha_{(1)}|}\sin^2(\pi \mu p)\, .
\EN
The difference from the scalar type II case treated 
in the previous subsection is of course the relative 
sign in the prefactor apart from the trivial difference  of 
the suppressed internal indices. By multiplying the 
factor $G$ which is the same as (\ref{correctionfactorg}), we confirm that 
the CFT coefficient (\ref{vscft1}) is precisely 
reproduced. 

\vspace{0.3cm}
In the present paper, 
we have restricted our discussions only to bosonic 
excitations. If we consider fermionic excitations, 
the correlation functions generically start from 
higher orders in $1/\mu$, since impurity non-preserving 
exchanges of BMN fermion fields with positive 
U(1)-charge $1/2$ between 2 and 3 in general require 
Yukawa-type interactions with scalar fields because 
of charge conservation. Furthermore, 
we have to take into account complex mixings 
among operators which have different numbers of 
bosonic and fermionic excitations but with degenerate conformal dimensions. For a precise treatment of these phenomena, it is  very 
important to formulate the correspondence 
of supersymmetries between bulk and boundary. 
Supersymmetry would put various constraints on the 
correlation functions among both bosonic and fermionic 
impurities. For examples of such constraints 
in the context of the pp-wave limit, see {\it e.g.}
 \cite{beisert} 
\cite{gursoy} and references therein. 
We are planning to discuss these aspects separately 
in the next work.

\section{Remarks} 
\setcounter{equation}{0}
In this final section, we give a few remarks on some 
relevant issues related to the main problem of the present work. 
We hope that this is useful for having further perspective 
on our approach. 

\subsection{Uniqueness of the holographic relation 
and the integrability of string field theory}
The holographic relation summarized in section 2 
was derived on the basis of the GKP-Witten relation 
and its perturbative computation in the case of 3-point function using bulk-to-boundary propagator. 
The original effective action for supergravity modes on which our derivation is based is written in terms of a particular 
choice of fields \cite{lmrs} by which derivatives 
with respect to the AdS$_5$ spacetime coordinates 
are completely eliminated.  This is justified since 
the 3-point function 
is independent of field redefinition. 
To the present order of approximation, it is sufficient to 
show that a would-be interaction term which 
could be generated from the quadratic term 
by a field redefinition does not contribute 
to the 3-point function. For notational simplicity, 
let us take the case of a single scalar field $\phi$ with action
\[
\int d^5x \sqrt{g}\, {1\over 2}[(\partial \phi)^2 
+ m^2\phi^2]\, .
\]
By a field redefinition 
$
\phi \rightarrow \phi + c \phi^2 ,
$
a 3-point interaction vertex 
\[
v_3=\sqrt{g}\, c\phi^2 ({\cal K}\phi), \quad {\cal K}=-{1\over \sqrt{g}}\partial_i \sqrt{g} \partial _i +m^2
\]
is generated. However, since the bulk-boundary propagator 
$\Delta(z, \vec{x}; \vec{y})$ satisfies 
the equation of motion ${\cal K}\Delta(z, \vec{x}; \vec{y})=0$ 
for arbitrary bulk spacetime positions $(z, \vec{x})$, 
we conclude that this vertex $v_3$ does not 
contribute to the correlation function, 
according to the GKP-Witten prescription. 
Therefore, the holographic relation before the 
large-$J$ limit is unique. 

Actually, in the special case of the so-called 
extremal correlators 
with $\Delta_2+\Delta_3-\Delta_1=0$, there is a well known 
subtlety. For generic configurations of U(1) $R$-charge, this is possible only for protected supergravity modes. The on-shell matrix elements of 
3-point interaction vertex in the bulk vanish for this case, 
but the 3-point correlation functions do not. 
In our formalism,   
the zero of the 
interaction vertex is canceled by the zero of the 
denominator in relating the interaction vertex to 
the CFT coefficient. In this sense, the extremal 
correlator has the $0/0$ ambiguity. However, we can 
circumvent this problem in the GKP-Witten relation 
and in our holographic relation 
if we slightly shift the conformal dimensions 
$\Delta_i \rightarrow \Delta_i +\epsilon_i$ by analytic 
continuation using the generic 
matrix elements of the interaction vertex such that 
the degeneracy $\Delta_2+\Delta_3=\Delta_1$ 
is lifted and, 
take the limit $\epsilon_i\rightarrow 0$ finally 
after we 
compute the CFT coefficients.

By taking the large-$J$ limit, we arrived at the 
first order action described in section 2 which describes 
the dynamics along a single tunneling trajectory. A question now 
arises as to the possibility of similar field redefinitions 
after the large-$J$ limit. Here we have to be very careful: 
In computing 3-point correlation functions using 
the tunneling trajectory, we had to 
introduce a cutoff near the boundary, since 
the approximation of a single trajectory is 
violated near the boundary even if we take 
a short distance limit by which two among three boundary 
points approach to each other up to a 
short distance of order $\delta$. 
Finding the precise prescription for this cutoff was 
the essence of section 3 of our previous work \cite{dy}.\footnote{
To avoid possible confusions, we emphasize that taking the short-distance limit $\delta\rightarrow 0$ here is 
different from considering double-trace operators. 
In our previous work \cite{dy}, 
this limit was used for the purpose of extracting the 3-point OPE 
coefficients. Its connection with the mixing of 
double-trace operators for the impurity-preserving sector 
was clarified there. For impurity non-preserving processes, 
3-point functions cannot be related to the mixing 
matrix of a dilatation operator, since by definition 
we are treating operators 
with {\it definite} conformal dimensions which are eigenstates 
of dilatation.}
Namely, the integral with respect to the
positions of interacting point along the tunneling trajectory 
in the naive perturbation theory 
\[
\int_{-\infty}^{+\infty}d \tau \, 
\e^{-(\Delta_1-\Delta_2 -\Delta_3)\tau}
\]
must be replaced by 
\EQ
\int_{-\infty}^{+\infty} d\tau \, 
\e^{-(\Delta_1-\Delta_2-\Delta_3)\tau}
\exp[-f{J_2J_3\over  J_1}{(2\delta)^2\over 
|\vec{x}_1-\vec{x}_c|^2}\e^{2\tau}]
\label{taucutoff}
\EN
where $f$ plays the role of taking into account the 
$\alpha'$ corrections. This implies that we cannot 
ignore  total derivatives of the type 
\[
{d \over d\tau}\e^{-(\Delta_1-\Delta_2 -\Delta_3)\tau}
=-(\Delta_1-\Delta_2 -\Delta_3)\e^{-(\Delta_1-\Delta_2 -\Delta_3)\tau}
\]
inside the integral.  It should be remarked that 
for this integral to be well defined we have to assume an 
analytic continuation in $\Delta_1-\Delta_2-\Delta_3$.  
 Thus the shift 
$\Delta_i\rightarrow \Delta_i +\epsilon_i$ of 
conformal dimensions which
is necessary for the extremal case is actually 
a general premise for our approach. 
Furthermore, the short distance limit $\delta\rightarrow 0$ must be 
taken after evaluating the integral 
with $f{J_2J_3\over  J_1}\delta^2=
fy(1-y)J_1\delta^2$ being kept fixed. 
In other words, the order of two limits,  
large $J$ and small $\delta$,  must be carefully 
chosen such that the integral is meaningful. 
Note that by overall scaling $\vec{x}_r 
\rightarrow \vec{x}_r/\delta$ the {\it short}-distance 
limit can equivalently be regarded as a 
{\it large}-distance limit in which $|\vec{x}_{23}|$ is 
fixed while $\vec{x}_1$ is sent to infinity. 

The expression (\ref{taucutoff}) is basically why we cannot 
impose conservation of energy for our `Euclidean' S-matrix. 
 Recall that in the  argument 
for the  uniqueness of 3-point functions before 
the large-$J$ limit, total spacetime 
derivatives are assumed to be vanishing. 
That is justified in the case of the GKP-Witten relation since 
the integral with respect to the interaction point 
extends to the whole bulk spacetime. 
In contrast to this, in our tunneling picture emerging 
by taking the large-$J$ limit, we have to be very careful when total derivatives are involved.  

In connection with this, it is appropriate to reconsider the 
suggestion made in our first work \cite{dsy} concerning a possible 
integrable structure of holographic string field theory, namely, the 
interaction term is actually obtained from the free theory 
by making a similarity transformation expressed using the CFT 
coefficient. In the present first order approximation, 
this can equivalently be formulated by a canonical transformation
\EQ
\psi(\tau) \rightarrow \psi_c(\tau) =\psi(\tau) + {1\over 2}\tilde{C}
\psi^2(\tau) - \tilde{C}\overline{\psi}(\tau){\psi}(\tau)
\label{cano1}
\EN
\EQ
\overline{\psi}(\tau) \rightarrow \overline{\psi}_c(\tau)=
\overline{\psi}(\tau) -\tilde{C}\psi(\tau)\overline{\psi}(\tau)
+{1\over 2} \tilde{C}\overline{\psi}^2(\tau)
\label{cano2}
\EN
where $\tilde{C}$ is proportional to the CFT 
coefficient itself 
\EQ
\tilde{C}_{123}=\Big(f {J_2J_3\over J_1}\Big)^{(\Delta_2+\Delta_3-\Delta_1)/2}
{C_{123}\over \Gamma({\Delta_2+\Delta_3-\Delta_1 \over 2}+1)}\, 
\EN
because of the relation between the CFT coefficient 
and the 3-point coupling $\lambda_{123}$
\EQ
(H^{(2)}_2+H^{(3)}_2-H^{(1)}_2)\tilde{C}_{123}
=\lambda_{123}\, . 
\EN
Note that we are here using a symbolic notation 
for brevity. 
It should be a trivial task to convert those formal 
expressions  
using the bra-ket notation for string fields. 
As is easily seen, the canonical transformation, (\ref{cano1}) 
and (\ref{cano2}), generates the 3-point 
interaction term $\lambda_{123}$ from the free Hamiltonian. 
This shows that the equation of motion of our holographic 
string field theory is {\it integrable} provided that the CFT coefficient is 
well defined. 

Does this imply that the three-point functions are 
obtained from the free string-field theory by the field redefinition 
corresponding to this canonical transformation? 
The answer to this question is clearly no, 
since the Euclidean S-matrix is invariant under field redefinition. 
Recall that the equivalence with respect to the 
equation of motion does not in general imply the 
equivalence of quantum mechanical amplitudes, because 
the surface terms often play crucial roles. 

Let us briefly check this 
using the LSZ formalism.  The 3-point 
S-matrix elements are defined as the residue 
of the 3-point Green functions 
$\langle \psi(\tau_1) \overline{\psi}(\tau_2)
\overline{\psi}(\tau_3)\rangle $ at the poles at 
$\omega_1=-i\mu\Delta_1, \omega_2=i\mu \Delta_2$ and 
$\omega_3=i\mu\Delta_3$, respectively, in the complex-energy ($\omega_i , \, 
i=1,2,3$) 
plane corresponding to external lines. If the field redefinition 
could generate a 3-point S-matrix element from the free theory, 
that would 
be obtained from the Green function 
\EQ
\langle \psi_c(\tau_1) \overline{\psi}_c(\tau_2)
\overline{\psi}_c(\tau_3)\rangle_{free}
\EN
where we compute the Green function using the free field action 
for the original string fields $\psi(\tau)$ and $\overline{\psi}(\tau)$ 
(not for the transformed fields $\psi_c$ and $\overline{\psi}_c$). 
Assuming that the $R$-charge angular momenta of external states 
are $J_1 \, (=J_2+J_3)$, $J_2$ and $J_3$ respectively 
corresponding to our convention throughout the present work, 
it is sufficient to 
study the following part of this Green function,
\EQ
\langle \psi_c(\tau_1) \overline{\psi}(\tau_2)
\overline{\psi}(\tau_3)\rangle_{free}
\sim {1\over 2}\langle \tilde{C}\psi^2(\tau_1) \overline{\psi}(\tau_2)
\overline{\psi}(\tau_3)\rangle_{free}
=\tilde{C}\langle \psi(\tau_1)\overline{\psi}(\tau_2)\rangle_{free}
\langle \psi(\tau_1)\overline{\psi}(\tau_3)\rangle_{free}\, .
\EN
Note that in this computation we do not use any partial integration with respect to the 
positions of interaction, 
since we do not make the change of field variables  explicitly for the action itself. 
This expression has poles at the correct positions in the complex energy plane 
for the lines 2 and 3, but 
 {\it not} for the line 1 \footnote{
The cutoff (\ref{taucutoff}) does not 
affect the positions of the leg poles. 
} 
 except for the extremal case where the interaction vertex itself vanishes. As above, 
we assume the prescription $\Delta_i\rightarrow \Delta_i 
+\epsilon_i$ for treating 
the extremal correlators
  to remove the degeneracy.  Under this definition, 
the Green functions in general, 
including the extremal case, do not have correct 
poles corresponding to external line 1, and 
hence the S-matrix elements vanish. 
Therefore, the free action 
cannot in general generate the S-matrix elements. 
\begin{center}
\begin{figure}
\begin{picture}(100,60)
\put(160,0){
\includegraphics[width=120pt]{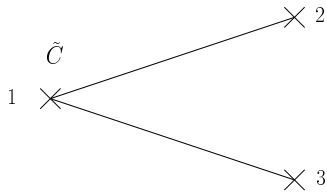}
}
 \end{picture}
\caption{The diagram which contributes to the Green function of free 
string field after the field redefinition.  }
\label{fig5} 
\end{figure}
\end{center}

\vspace{-1.3cm}
When we represent the Green function 
in the path-integral formalism, we would 
be able to discuss the above 
problem by making the change of the 
integrated string-fields from $(\psi,\overline{\psi})$  
to the redefined ones $(\psi_c, \overline{\psi}_c)$. 
Then, it would transform the free action to 
one with the interaction term and with crucial 
surface terms. The above direct 
treatment without using the change of 
the integration variables clearly shows that 
such an action generated by 
the canonical transformation 
does give only vanishing 
S-matrix, due to cancellation between 
local interaction terms and surface terms 
which are simultaneously generated by the 
change of integration variables.  To summarize, 
the conclusion of this subsection is that our holographic relation 
is essentially unique, independently of possible field redefinitions, 
from both viewpoints of the large-$J$ limit of the GKP-Witten relation and 
of the logic of string field theory to the present order of 
approximations.

In passing, we note that 
the above ambiguity of the extremal case in the bulk 
S-matrix is essentially the same thing on the gauge-theory side 
as the ambiguity 
mentioned in \cite{Kristjansen et. al.} with respect to 
the mixing of sugra BMN operators with double-trace operators.  
 As we have argued in \cite{dy}, however, 
we should not mix the double-trace operators for sugra modes.  
But if one wishes, one could mix 
the double-trace operators and correspondingly 
make a field-redefinition appropriately such that 
one obtains the same 3-point extremal 
correlators on both sides. The degree 
of arbitrariness is the same on both sides. 
So the existence of this ambiguity 
itself may be regarded as the consistency 
of our Euclidean S-matrix interpretation of 
holographic mapping in the PP-wave limit. 
We do not however adopt this viewpoint since 
without analytic continuation the extremal correlators 
cannot be computed from string-field theory umambigously.  
This peculiar ambiguity is related 
to a well known accidental degeneracy 
among multi-particle states  of massless particles  
with strictly collinear momenta.   
In any case, there is no ambiguity at all 
for impurity non-preserving 
cases treated in the present paper.

\subsection{Higher-order effects}
Extension of our ideas to higher orders is 
an important next problem. 
In the present paper, we have only treated the 
leading order results with respect to 
$1/\mu$-expansion (more precisely, $1/\mu\alpha_{(r)}\sim 
R^2/J$-expansion) to the 
first order in the genus-expansion parameter 
$1/N$. But in principle our holographic 
relation should be valid to all order in 
$1/\mu$ within the planar approximation. 
Actually, it is straightforward to 
compute next or next-to-next orders 
on the string side, since the Neumann functions 
are known \cite{He et. al.}\cite{Lucietteetal} apart from nonperturbative contributions.  
Corresponding computations on the gauge-theory 
side, however,  require us to obtain 3-point functions 
at least up to 2-loop orders, including all 
mixing effects. 
Once we go to such higher-orders, 
an interesting puzzle arises.  The factor $G$ which 
played crucial roles in the foregoing discussions 
gives  corrections of the form  
$\mu^{-2n}\log^m \mu \, \, (n>0)$, since the 
large $\mu$ behavior (\ref{asympf}) of the function $f$ 
is correct up to nonperturbative exponential corrections, 
according to the above references. 
If we ignore the would-be nonperturbative 
terms of the Neumann functions, this implies that,  
when they are 
not protected, 3-point correlation functions 
 are subject to these peculiar 
corrections which cannot be derived 
by the ordinary perturbation theory on the
gauge-theory side.  

At present, we are not 
sure how this should be interpreted. A possibility may be 
that the total sum of nonperturbative corrections 
on the Neumann functions  might generate 
such logarithmic terms and cancel 
these corrections when $m\ne 0$. For possible 
subtleties of  nonperturbative corrections, we refer the reader to  \cite{Kleba} and also to 
\cite{Lucietteetal}. Or, if such 
terms with nonzero $m$ 
are not canceled on the string side,  
nonperturbative 
effects (or exact summation over the whole 
perturbation series) both on the gauge-theory side and on the 
string side might be 
responsible for complete understanding of 
string/gauge duality. Of course, it is also possible that 
our holographic relation itself may be subject to 
corrections. 

The higher orders with respect to $1/N$ is also 
important. Since our holographic relation 
is originally derived for sugra modes in the 
planar approximation, it is not immediately 
clear whether our conjecture should be 
valid for nonplanar cases. 
In principle, we have to examine the same limiting 
procedure for such cases as we have discussed 
for planar 3-point functions. 
Following our philosophy, all rules of the holographic 
mapping should be derived on the basis of 
the correspondence between correlation functions 
on both sides. As for higher-genus corrections, 
it is not evident 
whether the computation of loop corrections 
on the string side can be closed within the 
PP-wave in the presence of the sum over 
intermediate states. Recently, it has been reported \cite{Gutpan} that 
the energy shift computed from the 
$|H_3\rangle_{SV}$ -type string vertex 
disagrees with the gauge-theory prediction even 
at the leading order in the first non-planar correction, if 
the intermediate states are restricted to impurity-preserving 
sector.

Even if it could be closed 
within the PP-wave limit, we have to fix 
possible higher-order corrections for 
susy generators and Hamiltonians themselves on the 
string field side. In our first paper \cite{dsy} 
in this series, we have suggested that  in planar (tree in the sense of string field theory) approximation, the higher-point vertices may be 
derived by a similarity transformation from free theory whose 
lowest-order form is nothing but the 
canonical transformation discussed 
in the previous subsection. If we include non-planar 
corrections, these (similarity or perhaps unitary in the 
space of string fields) transformations 
generate $1/N$ corrections  for vertices and 
even for quadratic kinetic terms.  In order to confirm this suggestion, 
it is important to investigate first the extension of our 
arguments for higher-point amplitudes in the planar limit. 
It has been pointed out that the structure of 4-point correlation 
functions on the gauge-theory side suffers from some 
ambiguity \cite{Kristjansen et. al.} 
in determining the large-$J$ limit. 
    From our point of view,  we have to take into account 
the fact that the approximation of a single tunneling 
trajectory can be justified only if we first take the 
short-distance limit for operator products such that 
the  order $\delta$ of the distances at each  set of 
operator products at initial and final points at the boundary 
 satisfies $\delta^2 \sim c/J$ with some finite $c$ and 
finally let $c\rightarrow 0$. 
It is clear that the 
naive limit $J\rightarrow \infty$ on fixing the space-time 
configuration of operators at the boundary,  which indeed 
leads to discontinuous results depending on the 
positions of  operators, is not allowed. 
Careful analyses are required, however,  to make 
more precise the general structure 
suggested in \cite{dsy} for 
holographic mapping  for higher-point 
S-matrix elements. 

{\it Note added}: After completing this manuscritpt, 
we received a preliminary manuscript by H. Shimada \cite{shima} 
who tried to establish the holographic relation in an approach 
which is slightly different from ours. His argument and result, 
though restricted yet to impurity-preserving sector with 
scalar excitations, seem to be consistent 
with ours in \cite{dy} and complementary to our argument 
 for the uniqueness of the holographic relation. 
 
\vspace{0.5cm}
\noindent
{\large Acknowledgements}

We would like to thank H. Shimada for discussions. 
The present work is supported in part 
by Grant-in-Aid for Scientific Research (No. 13135205 (Priority Areas) and No. 16340067 (B))  
from the Ministry of  Education,
Science and Culture. 

\appendix 
\section{Large $\mu$ behavior of Neumann coefficients}
\setcounter{equation}{0}
\renewcommand{\theequation}{\Alph{section}.\arabic{equation}}
\renewcommand{\thesubsection}{\Alph{section}.\arabic{subsection}}

We summarize here 
the leading large $\mu$ behavior of $\widetilde N_{mn}^{rs}$ 
for reader's convenience following \cite{He et. al.}, 
\cite{Lucietteetal} :
\begin{eqnarray}\label{Neumann for large mu}
&&\hspace{-1cm}\widetilde N_{mn}^{\inI\inI}=\frac{(-1)^{m+n}}{4\pi\mu |\aout|y },
\quad
 \widetilde N_{mn}^{\inI\inII}=\frac{(-1)^{m+1}}
{4\pi\mu|\aout| \sqrt{y(1-y)}}\\
&&\hspace{-1cm}\widetilde N_{mn}^{\inII\inII}=\frac{1}{4\pi\mu |\aout|(1-y)},
\quad
 \widetilde N_{mn}^{\out\out}=\frac{(-1)^{m+n+1}\sin(\pi m y)\sin(\pi n y)}
{\pi \mu|\aout|}\\
&&\hspace{-1cm} \widetilde N_{mn}^{\inI\out}
=\frac{(-1)^{m+n+1}\sin(\pi n y)}
{\pi \sqrt{y}(n-m/y)},\quad
\widetilde N_{mn}^{\inII\out}=\frac{(-1)^{n}\sin(\pi n y)}
{\pi \sqrt{1-y}(n-m/(1-y))}
\end{eqnarray}
for $(m,n)\neq(0,0)$, and 
\begin{eqnarray}
&&\hspace{-1cm}\widetilde N_{00}^{\out\out}=0,\quad
 \widetilde N_{00}^{\out\inI}=-\sqrt{y},\quad
 \widetilde N_{00}^{\out\inII}=-\sqrt{1-y},\\
&&\hspace{-1cm}\widetilde N_{00}^{\inI\inII}
=-\frac{1}{4\pi\mu|\aout| \sqrt{y(1-y)}},
\quad
\widetilde N_{00}^{\inI\inI}=\frac{1}{4\pi\mu|\aout| y},
\quad
 \widetilde N_{00}^{\inII\inII}=\frac{1}{4\pi\mu|\aout| (1-y)},
\end{eqnarray}
for $m=0=n$.

The asymptotic form in the large $\mu$ limit of $f=1-4\mu \a_{(1)}\a_{(2)}\a_{(3)}K$ is given by
\begin{eqnarray}\label{asymptotic forms}
&&\hspace{-1.5cm} f= \frac{1}{4\pi \mu |\aout| y(1-y)}. \nonumber 
\end{eqnarray}

\end{document}